\documentclass[aps,prl,reprint,superscriptaddress]{revtex4-1}
\usepackage{hyperref}
\usepackage{braket}
\usepackage{graphicx}
\usepackage[caption=false]{subfig}
\begin{document}
\title{Revealing the nonlinear response of a tunneling two-level system ensemble using coupled modes} 
\author{Naftali Kirsh}
\affiliation{Racah Institute of Physics, the Hebrew University of Jerusalem, Jerusalem, 91904 Israel}
\author{Elisha Svetitsky}
\affiliation{Racah Institute of Physics, the Hebrew University of Jerusalem, Jerusalem, 91904 Israel}
\author{Alexander L. Burin}
\affiliation{Department of Chemistry, Tulane University, New Orleans, Louisiana 70118, USA}
\author{Moshe Schechter}
\affiliation{Department of Physics, Ben-Gurion University of the Negev, Beer Sheva 84105, Israel}
\author{Nadav Katz}
\email{katzn@phys.huji.ac.il}
\affiliation{Racah Institute of Physics, the Hebrew University of Jerusalem, Jerusalem, 91904 Israel}
\date{\today}

\begin{abstract}           
Atomic sized two-level systems (TLSs) in amorphous dielectrics are known as a major source of loss in superconducting devices. In addition, individual TLS are known to induce large frequency shifts due to strong coupling to the devices. However, in the presence of a broad ensemble of TLSs these shifts are symmetrically canceled out and not observed in a typical single-tone spectroscopy experiment. We introduce a two-tone spectroscopy on the normal modes of a  pair of coupled superconducting coplanar waveguide resonators to reveal this effect. Together with an appropriate saturation model this enables us to extract the average single-photon Rabi frequency of dominant TLSs to be $\Omega_0/2\pi \approx 79 $ kHz. At high photon numbers we observe an enhanced frequency shift due to nonlinear kinetic inductance when using the two-tone method and estimate the value of the nonlinear coefficient as $K/2\pi \approx -1\times 10^{-4}$ Hz/photon. Furthermore, the life-time of each resonance can be controlled (increased) by pumping of the other mode as demonstrated both experimentally and theoretically. 
\end{abstract}

\maketitle

The characterization of a nonlinear medium often involves the usage of a strong (pump) field which modifies the medium properties, combined with a weak (probe) field which is used for measurement. When the nonlinear response of a resonant device is probed, the finite linewidth limits the available pumping bandwidth  
and thus might hide the full nonlinear behavior. Here we show how this problem can be solved by using the normal modes of a coupled system. 
When two modes share the same spatial volume but have different frequencies, one mode can be pumped strongly, modifying the medium locally overlapping with the other mode which is used for probing. When the splitting between the modes is large enough the pumped mode can modify the spectral components of the medium outside of the other mode's linewidth. In this Rapid Communication we demonstrate how this method can reveal nonlinear properties which would have remained hidden even for the strongest drive possible if the standard method was used. In particular, we measure low-power nonlinear frequency shifts of resonances formed by coupled superconducting coplanar waveguide resonators (CPWR) \cite{GoppelCPW}. As we show, these shifts, caused by pumping at a detuning of $\sim 1000$ times the resonance linewidth, 
are well explained by the saturation of two-level systems (TLSs) in the device's dielectrics. 
Thus, this method can be used to give valuable information about TLSs, which was unavailable otherwise.    
Furthermore, at high powers, when kinetic inductance nonlinearity is dominant, our two-tone method provides twice the nonlinear sensitivity, which might be useful for applications such as microwave kinetic inductance detectors \cite{MKID} and resonators' frequency tuning \cite{KI_tuning}.

Investigated in the context of amorphous material physics \cite{Anderson,phillips1972tunneling,Phillips}, TLSs in dielectrics were brought into focus again since the discovery of their critical role as a loss mechanism in superconducting devices \cite{MartinisTLS}. While the effect of TLSs saturation by probe power on the imaginary part of the dielectric constant $ Im\left(\epsilon\right)$ (i.e. internal loss) was investigated thoroughly \cite{standardLoss,MartinisTLS,PappasTLS,KhalilTLSloss,StudyOfLoss,rosen2016random,skacel2015probing}, the modification of $ Re\left(\epsilon\right) $ (i.e. frequency shift) by the drive has remained hidden \cite{TempNote}. Here we demonstrate how  by applying a pump-probe
  measurement on coupled modes this effect can be revealed. 

\begin{figure}[h!]
	\includegraphics[trim={4.5cm 0cm 4.4cm 0cm},clip,width=0.8\linewidth]{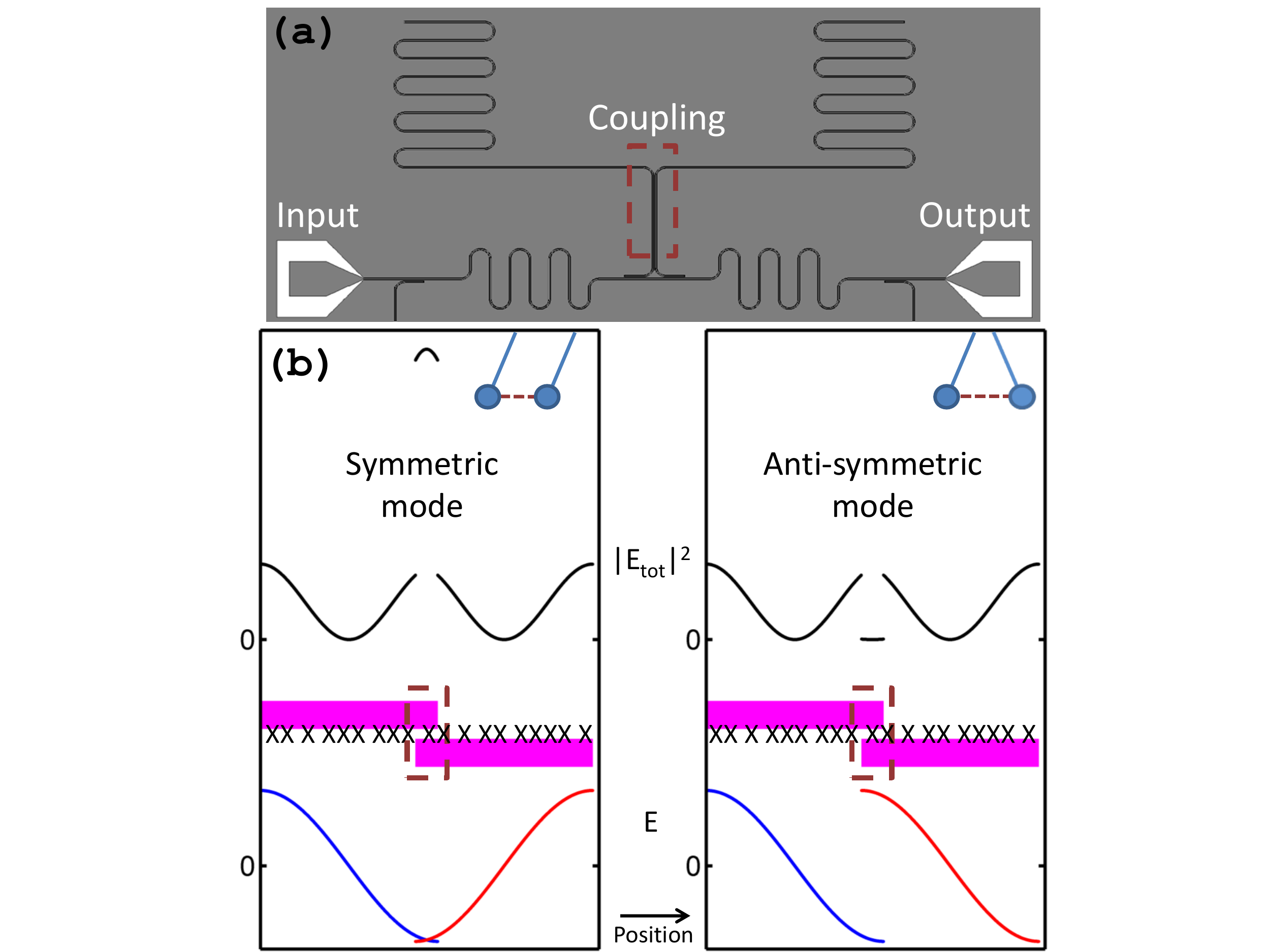} 
	\caption{(a) Design of the coupled resonators device \cite{ResNote}. Empty area represents Si substrate, gray area represents metal (Al). Feedline and resonators are black for clarity. Dark-red dashed box marks the coupling region. (b) Cartoon of the experimental idea. Magenta bars represent the resonators, the $x$ axis is the position along their lengths. In analogy to coupled pendula the coupling results in the forming of a symmetric and an antisymmetric modes which differ only by the phase between the electric fields in both resonators (blue and red curves) \cite{PendulaNote}. TLSs (represented by 'X') experience the same total field intensity (black curves) in both modes, except for the small coupling area in which the resonators' fields interfere. This allows ignoring other details than the frequency detuning between the modes.}
	\label{fig:device}
\end{figure}

Coupled resonators with a strong Josephson nonlinearity were used e.g. for quantum-limited amplification \cite{WallraffCoupled}, stabilization of photon-number states \cite{Holland2015} and for simulating a Bose-Hubbard chain \cite{BoseHubbard}.  
Here we use coupled resonators as a tool to characterize the intrinsic nonlinearities of the resonators. For an ideal coupled pair of two identical resonators the normal modes will be a symmetric mode and an antisymmetric one for which the wave functions are identical up to a phase. This allows us to ignore uncertainties e.g. in electric field distribution and treat the modes referring only to the ability to saturate detuned spectral components of the nonlinear medium (see Fig.~\ref{fig:device}).           

Our device consists of two capacitively 
coupled $\lambda/2$ CPWR made from $\sim120$ nm Al sputtered on high-resistivity Silicon ($\rho>5000\,\,\Omega\cdot$cm), see Fig.~\ref{fig:device}.
These resonators are physically identical, hence their bare (uncoupled) resonance frequencies are the same, but due to the coupling, two modes split by $\sim 63$ MHz at 5626 and 5689 MHz are formed. 
We characterize the resonances by measuring the transmission through a common feedline to which they are capacitively coupled using a vector network analyzer. The resonance frequency $f_0$, internal quality factor $Q_i$ and the steady-state average number of photons $\braket{N}$ stored in the resonator are extracted by fitting the complex transmission data $S_{21}\left(f\right)$ to an appropriate model \cite{Khalil,MazinThesis,UstinovMethod,SuppNote}. In order to quantify the nonlinear response of the resonances the probe power is scanned at a range which corresponds to $\langle N \rangle = 10^1-10^8$ photons, and the dependence of various parameters on $ \langle N \rangle $ is analyzed. All measurements were conducted at a temperature of $\sim 20$ mK with the device mounted to the base plate of a dilution refrigerator.  

\begin{figure}[h]
	\subfloat[]{
		\centering
		\includegraphics[trim={1.5cm 3.2cm 2cm 0cm},clip,width=\linewidth]{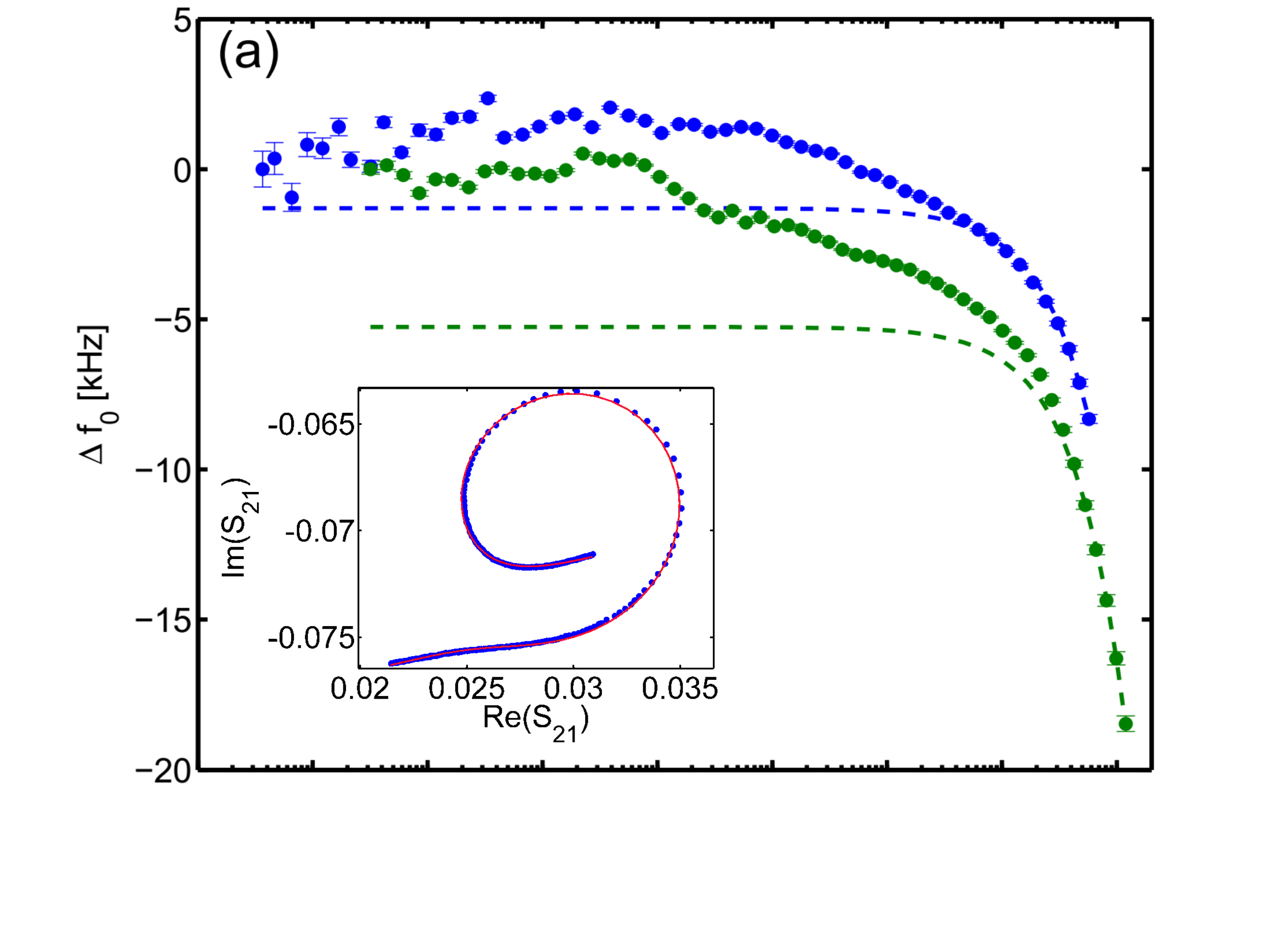} 
		\label{fig:oneToneF0}
	}
	\hfill
	\subfloat[]{
		\centering
		\includegraphics[trim={1.5cm 1.2cm 2cm 0cm},clip,width=\linewidth]{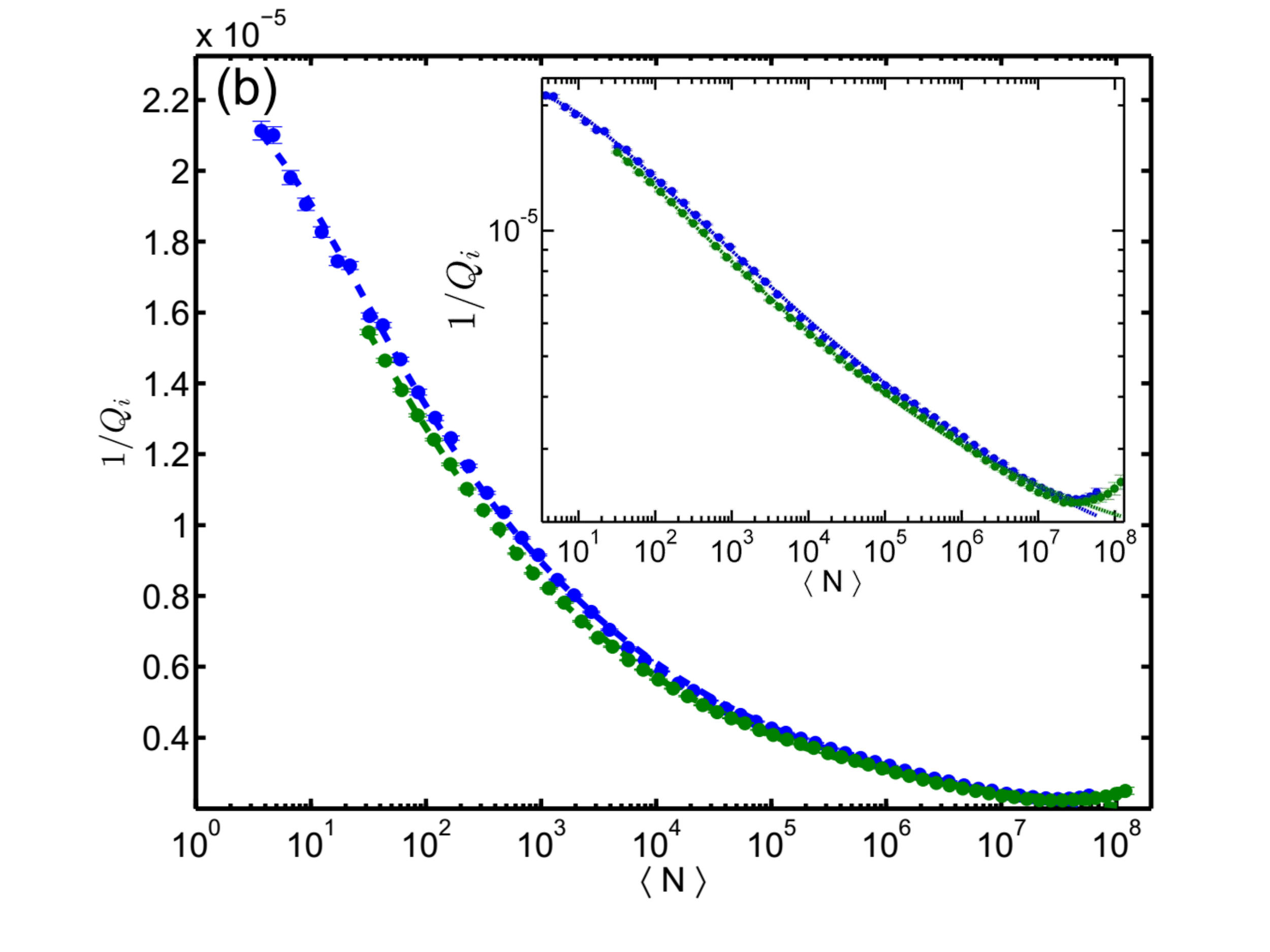} 
		\label{fig:oneToneLoss}
	}
	\caption{(a) Resonance frequency shift and (b) Internal loss (shown in a log-log scale in the inset) vs $\langle N \rangle$ for the probe-only experiment for the resonances at 5626 MHz (green) and 5689 MHz (blue). Dashed lines in \ref{fig:oneToneF0} are linear fits of the high power shifts to a Kerr nonlinearity model as detailed in the text. Dashed lines in \ref{fig:oneToneLoss} are fits to the standard TLS model (\cite{SuppNote} Eq.~(29)). Inset in Fig.~\ref{fig:oneToneF0}: a typical fit of the transmission data.}
	\label{fig:oneTone}
\end{figure}

\begin{figure}
	\includegraphics[trim={0.5cm 1.5cm 0cm 2.5cm},clip,width=\linewidth]{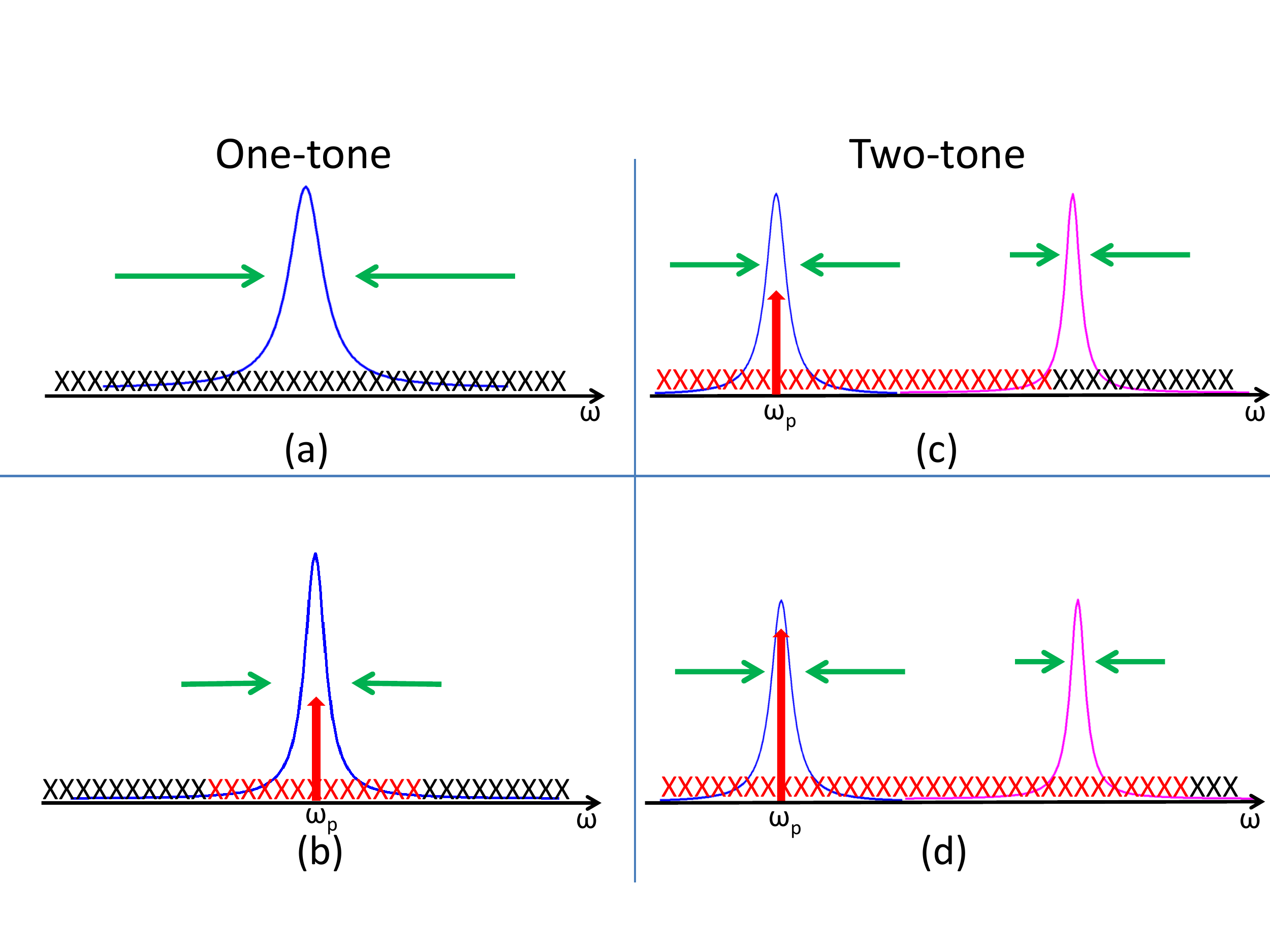} 
	\caption{Qualitative picture of the generalized TLS saturation model for the resonance shift. A
		black 'X' represents a TLS. A red 'X' represents a saturated TLS. Green
		arrows represent the level repulsion (the length indicating the repulsion strength). Red
		arrows represent the pump tone. (a) In the absence of pumping the TLSs are
		symmetrically distributed around a resonance, resulting in a broadening without a frequency shift. 
		(b) A symmetric pump around one resonance narrows the linewidth without a
		frequency shift. (c) When two resonances are present, strongly pumping
		symmetrically around the lower resonance results in an asymmetric repulsion for the higher
		resonance, causing a frequency shift. (d) When the pump power increases, the 
		repulsion of the higher resonance becomes more and more symmetric, causing a
		decrease in the frequency shift.}
	\label{fig:qualitative}
\end{figure}

In Fig.~\ref{fig:oneTone} we show the resonance frequency $ f_0 $ and internal loss $1/Q_i$ of the resonances obtained using the standard single-tone spectroscopy method (for a uniform medium and electric field $1/Q_i=\tan \delta$, the bulk loss tangent, see~\cite{SuppNote}).  
 The frequency of a single-tone probe is scanned around one resonance for various probe powers and the parameters are extracted as detailed above. As shown in Fig.~\ref{fig:oneToneLoss} the loss dependence on $ \langle N \rangle $ agrees well with the TLS model \cite{SuppNote}. In addition, as predicted by this model \cite{PappasTLS}, for stored energies corresponding to $\langle N \rangle \lesssim  10^5$ there is no observable frequency shift. For higher probe powers there is a strong negative shift which depends linearly on the number of photons. This linear shift is explained as resulting from nonlinear kinetic inductance \cite{ResonatorsReview} which can be modeled as a Kerr nonlinearity \cite{YurkeBuks} (the weak negative shift for $10^5 \lesssim \langle N \rangle \lesssim 10^7$ resulting in a discrepancy from the fits at low powers might be due to the finite number of TLSs, see \cite{SuppNote}). 
We elaborate more on the high power regime below. We stress here that in fact TLSs far detuned from resonance contribute to the real part of the dielectric constant $ \epsilon $ \cite{PappasTLS} and therefore should effect $ f_0 $, but because the standard method uses a single tone which 
saturates TLSs symmetrically around resonance this nonlinear effect is effectively hidden (See Figs.~\hyperref[fig:qualitative]{4a} and ~\hyperref[fig:qualitative]{4b}).  

\begin{figure}[h]
	\subfloat[]{
		\centering
		\includegraphics[trim={1cm 7.8cm 1cm 9cm},clip,width=\linewidth]{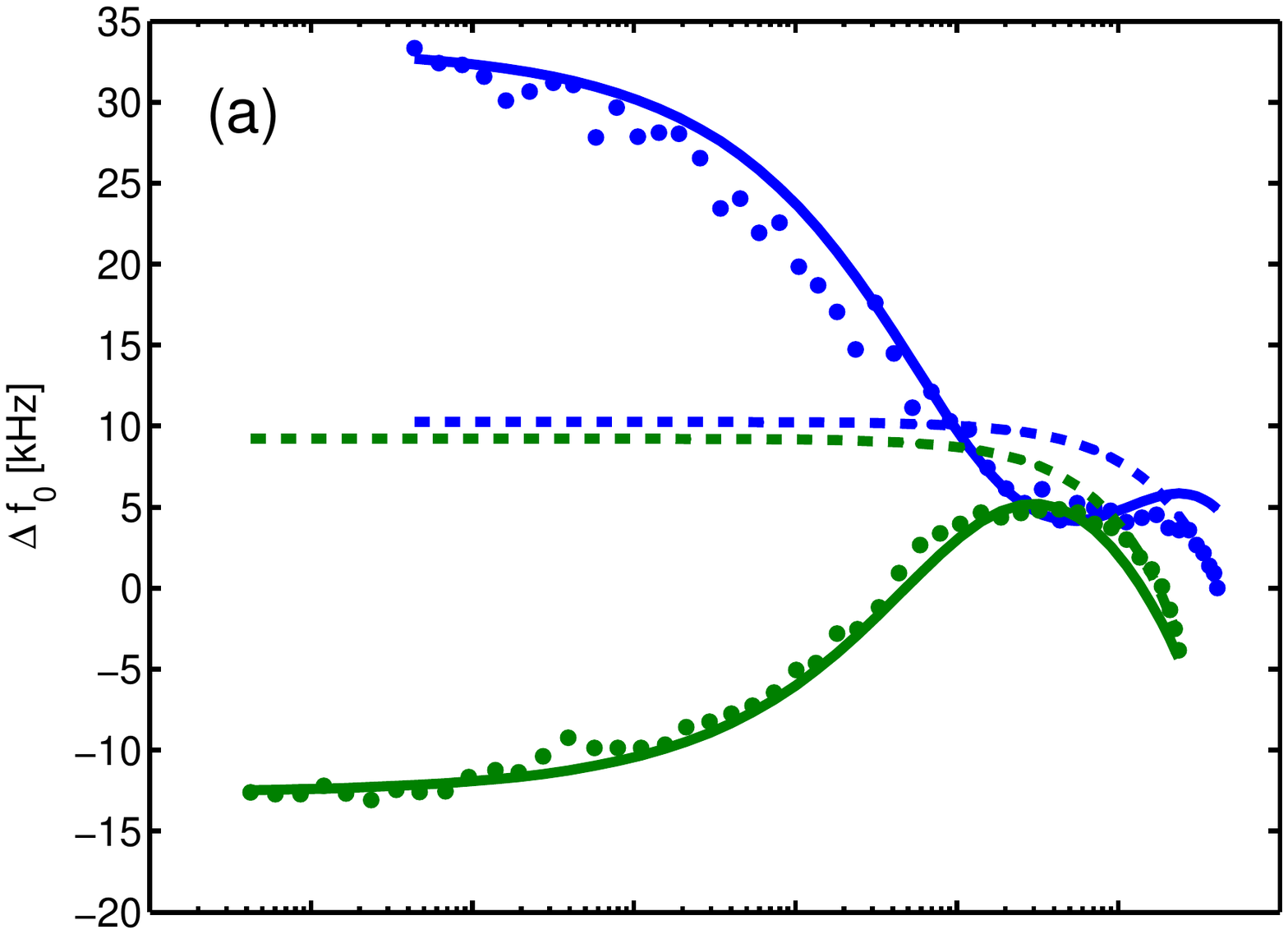} 
		\label{fig:deltaF0TwoTone}
	}
	\hspace{0cm}
	\subfloat[]{
		\centering
		\includegraphics[trim={1cm 6cm 1cm 8.4cm},clip,width=\linewidth]{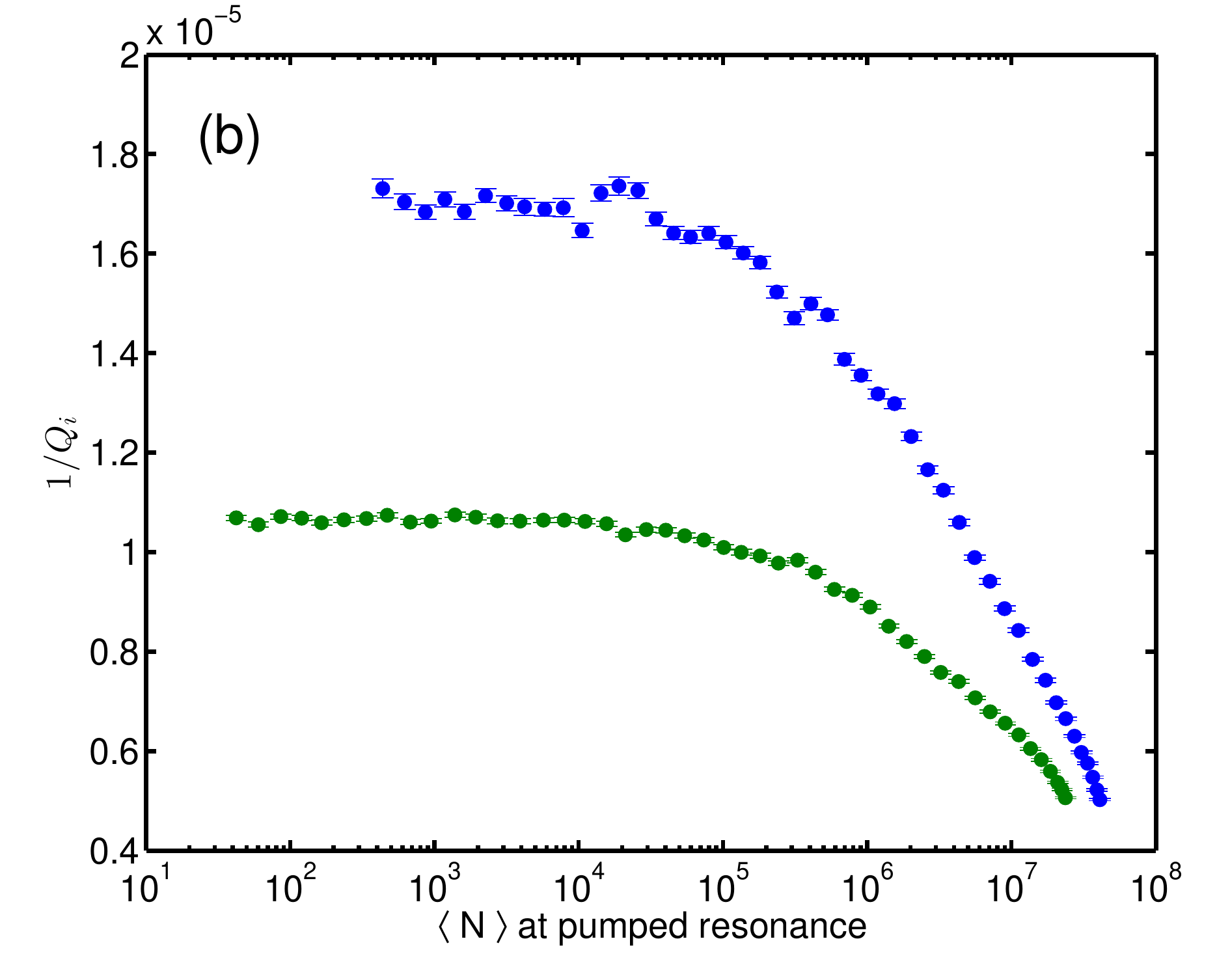} 
		\label{fig:lossTwoTone}
	}
	\caption{(a) Resonance frequency shift and (b) Internal loss vs the average number of photons $\langle N \rangle$ in the pumped mode for the pump-probe
		 experiment. Green (blue) points are for probe at 5626 MHz (5689 MHz) and pump at 5689 MHz (5626 MHz). The dashed lines in Fig.~\ref{fig:deltaF0TwoTone} are linear fits of the high power shifts to a Kerr nonlinearity model as detailed in the text. Solid lines are sums of the fits to the prediction of the generalized model of frequency shift due to TLSs (Eq.~(\ref{eq:f0Shift})) and the linear fits. In Fig.~\ref{fig:deltaF0TwoTone} constant shifts were added for clarity.}
	\label{fig:twoTone}
\end{figure}

The full nonlinear behavior of the resonances is revealed when we use our two-tone pump-probe method. This is done by pumping strongly at a frequency close to one resonance while measuring  $S_{21}\left(f\right)$ of  the other resonance using a weak probe. Fig.~\ref{fig:deltaF0TwoTone} shows the results of this experiment. Notice that while in Fig.~\ref{fig:oneTone} the $x$ axis indicates the average number of photons in the \textit{probed} resonance, in Fig.~\ref{fig:twoTone} the $x$ axis shows the average number of photons in the \textit{pumped} resonance and the weak probe power is held constant \cite{SuppNote}. In contrary to the kinetic inductance nonlinearity at high photon numbers which is common to both types of experiments, at lower powers there is a significant frequency shift in the pump-probe
 experiments which is not observed in the probe-only experiments. Furthermore, the direction of the resonance shift depends on the relative position of the pumped and probed resonances. A negatively-detuned pump induces a negative shift on the probed resonance and vice-versa. These shifts can be explained by generalizing the TLS model to the pump and probe case. We first give a qualitative explanation of the physical reasoning behind the differences and then give the details of the full derivation.    
The effect of one TLS on the probed mode frequency can be approximately described by the dispersive shift \cite{cQED} $ \Delta \omega_{pr} = 
\frac{g^2}{\Delta}\langle\hat{S}_z\rangle $, where $ g $ is the coupling strength between the TLS and the mode, $ \Delta \equiv \omega_{TLS}-\omega_{pr} $ is their detuning and $ \hat{S_z} = \pm 1 $ is a TLS state operator (this result is obtained by applying second order perturbation theory on the Jaynes-Cummings Hamiltonian \cite{JaynesCummings}, assuming $g<\Delta$ and can be understood as resulting from level repulsion between the mode and the TLS). Since TLSs density of states is uniform in energy \cite{Phillips} they will be distributed symmetrically around $ \omega_{pr} $ resulting in equal negative and positive shifts which sum to a zero shift with a broadened linewidth (Fig.~\hyperref[fig:qualitative]{4a}). When a TLS is pumped strongly it populates the excited state half of the time, yielding $\langle\hat{S_z}\rangle = 0 $. Nevertheless, when the standard probe-only method is used, TLSs are saturated symmetrically around $ \omega_{pr}$, hence the effect is only linewidth narrowing (i.e. reducing the internal loss) but no observable frequency shift (Fig.~\hyperref[fig:qualitative]{4b}). The situation is different when one mode is pumped while the other is probed. The presence of another mode opens a spectral window for asymmetric saturation of TLSs which effects the probed mode due to the spatial coexistence of both modes' fields. When a mode negatively (positivity) detuned from the probe field is pumped strongly TLSs negatively (positivity) detuned from $ \omega_{pr}  $ are saturated, while positivity (negatively) detuned TLSs are less affected, resulting in a negative (positive) net shift (Fig.~\hyperref[fig:qualitative]{4c}). Increasing pump power increases also the TLSs saturation (from power broadening), initially increasing the shift but eventually making saturation more and more symmetric, hence reducing the shift (Fig.~\hyperref[fig:qualitative]{4d}). We now present a full quantitative model to explain the results \cite{SuppNote}. A single TLS is characterized by an asymmetry energy $ \Delta_j $, a tunneling amplitude $ \Delta_{0j} $  resulting in the eigenenergy $ E_j \equiv \hbar\omega_j \equiv \sqrt{\Delta_j^2+\Delta_{0j}^2} $ and an electric dipole moment $ p_j $ \cite{Phillips}. The dipole coupling coefficient to a uniform electric field occupying a volume $ V $ filled by a material with a dielectric constant $ \epsilon $ is $ g_j \equiv \frac{\cos \theta_j p_j \sqrt{\omega_{pr}}}{\sqrt{2\hbar\epsilon\epsilon_0 V}} \frac{\Delta_{0j}}{\hbar\omega_j}$, where $ \theta_j $ is the angle between the electric field and the dipole and $\omega_{pr}$ is the probed mode frequency. Summing contributions from all TLSs we obtain the  
frequency shift       
\begin{equation}
\Delta\omega_{pr}=\mathrm{Re}\left[\sum_j g_j^2 \frac{\langle\hat{S}_{z,j}\rangle}{\omega_{j}-\omega_{pr}+\frac{i}{T_{2j}}}\right],
\end{equation}
where $ T_{2j}=2T_{1j} $ are the relaxation and dephasing times of the TLS (assuming a negligible TLS-TLS interaction). The population difference induced by pumping yields \cite{Phillips}
\begin{eqnarray}
\langle\hat{S}_{z,j}\rangle = -\tanh\left(\frac{\hbar\omega}{2k_BT}\right)\times \nonumber \\ 
\left(1-\frac{T_{1j}\Omega_{Rj}^2/T_{2j}}{\left(\omega_j-\omega_{pump}\right)^2+T_{2j}^{-2}\left(1+\Omega_{Rj}^2 T_{1j}T_{2j}\right)}\right)
\end{eqnarray}
where $ \omega_{pump} $ is the pump frequency and the TLS Rabi frequency due to pumping is given by
\begin{equation}
\Omega_{Rj} \equiv \frac{\Delta_{0j}}{\hbar\omega_j} \cos \theta_j \Omega_0,\,
\hbar\Omega_0 \equiv 2p_j F_{pump}, 
\end{equation}
with the pump field $ F_{pump} $. Using the standard TLS model distribution  
we obtain that in the strong field regime $ \Omega_0 T_1 T_2 \gg 1 $ the frequency shift due to pumping is \cite{SuppNote}
\begin{equation}
	\frac{\Delta\omega_{pr}}{\omega_{pr}\tanh\left(\frac{\hbar\omega_{pr}}{2k_BT}\right)} =  \frac{\sqrt{2}P_0 p^2 \pi^2}{8\epsilon\epsilon_0}\frac{\Delta\omega}{\Omega_0}\frac{\sqrt{1+\frac{\Omega_0^2}{2\Delta\omega^2}}-1}{\sqrt{1+\frac{\Omega_0^2}{2\Delta\omega^2}}+1},
	\label{eq:f0Shift}
\end{equation} 
where $ \Delta\omega \equiv \omega_{pump}-\omega_{pr}$ is the detuning between the pump and probe frequencies and $P_0$ is the TLSs density of states. As expected from the qualitative picture, for small fields $ \left(\Omega_0 \ll \Delta\omega\right) $ the shift increases with $ \langle N \rangle $ as $ \Delta\omega_{pr} \propto \Omega_0 \propto \sqrt{\langle N \rangle} $, while it decreases as  $ \Delta\omega_{pr} \propto \Omega_0^{-1} \propto \langle N \rangle^{-\frac{1}{2}} $ for high fields $ \left(\Omega_0 \gg \Delta\omega\right) $. In addition, the direction of the shift follows the sign of $ \Delta\omega $ \cite{SuppNote} as expected. In Fig.~\ref{fig:deltaF0TwoTone} we show fits to this model for both resonances \cite{SuppNote}. 
From the fits we obtain an estimate to the average single photon Rabi frequency of a dominant TLS $ \Omega_0^{N=1}/2\pi \approx 79 $ kHz \cite{SuppNote}. These values confirm the predictions of previous studies \cite{Xmon,MartinisSimulations} and of our Monte-Carlo simulations \cite{SuppNote} that TLSs at regions of strong fields dominate the nonlinear behavior. In comparison to the frequency shift the internal loss needs a much larger number of photons to show a significant response (Fig.~\ref{fig:lossTwoTone}), This agrees  with our theoretical calculations \cite{SuppNote} and demonstrates the fact that the imaginary part of $ \epsilon $ is only sensitive to nearly resonant TLSs \cite{PappasTLS}. 

As mentioned above, for large $ \langle N \rangle $ a negative frequency shift is observed in both probe-only and pump-probe
  experiments. This shift can be attributed to nonlinear kinetic inductance \cite{ResonatorsReview} which might be explained e.g. as resulting from quasiparticle microwave heating \cite{deVisserHeating} or modification of the superconducting ground state by the field \cite{coherentExcited}. Following Yurke and Buks \cite{YurkeBuks} we model this effect as a Kerr nonlinearity, resulting in a frequency shift which is linear in the number of photons
\begin{equation}
	\Delta f_0 = \frac{K}{2\pi}\langle N \rangle,
\end{equation}  
where $ K $ is the Kerr coefficient. 
A linear fit to the single-tone
 experiments high-power frequency shifts (Fig.~\ref{fig:oneToneF0}) yields $ \frac{K}{2\pi} = -1.11(0.05)\times10^{-4}$ 
Hz/photon  for the resonance at 5626 MHz and $ \frac{K}{2\pi} = -1.23\times10^{-4} $ 
Hz/photon for the resonance at 5689 MHz. These results agree with order of magnitude estimations based on a simple nonlinear kinetic inductance model \cite{SuppNote}. These close values for both resonances agree with the fact that kinetic inductance should depend on the superconducting Al properties \cite{DahmKineticInductance,deVisserHeating,coherentExcited}. Generalizing the model to the case of coupled resonators \cite{SuppNote} we find that the expected shift of one resonance due to photons in the other resonance is linear in the number of photons with the coefficient $ K^\prime=2K $, i.e. doubled. Fitting the two-tone experiments high-power frequency shifts (Fig.~\ref{fig:deltaF0TwoTone}) we obtain $ \frac{K^\prime}{2\pi} = -5(1)\times10^{-4}$  
Hz/photon for the 5626 MHz resonance and $ \frac{K^\prime}{2\pi} = -2.4(0.8)\times10^{-4}$
Hz/photon for the 5689 MHz resonance. For the 5689 MHz resonance we obtain an agreement with the expected theoretical doubling, but for the 5626 MHz resonance there is a discrepancy which might be a result of an additional TLS shift due to non-uniform electric field or imperfections in the calculation of the number of photons at high powers \cite{SuppNote}. 
Similar enhancement of the cross-Kerr shift in comparison to the self-Kerr one   
 was observed with a strong Josephson nonlinearity \cite{Holland2015}. Here it is used as a tool for measuring the intrinsic nonlinearity of a presumably linear resonator. 

In conclusion, a 
method for the characterization of nonlinearities using a pump-probe
 scheme on the normal modes of coupled resonators was implemented and analyzed. Using this method the effect of TLS saturation by drive power on the real part of the dielectric constant $ \epsilon $ was uncovered and the average single photon Rabi frequency of dominant TLSs was extracted. In addition, the Kerr coefficient quantifying the strength of nonlinear kinetic inductance was measured, yielding $ \frac{K}{2\pi} \approx -1\times 10^{-4} $ Hz/photon. Knowledge of the nonlinearities of presumably linear CPWRs down to the single photon limit is important for quantum information applications, such as the implementation of recent proposals for encoding logical qubits by multiphoton cat states \cite{cat}. In addition, the ability to reduce the internal loss of one mode by pumping the other mode can be used to enhance the resonance lifetime while keeping the number of probing photons small, as required e.g. for dispersive readout in circuit QED \cite{cQED}.  

\begin{acknowledgments}
	We thank Dr. Sebastian Probst and Prof. Lazar Friedland for fruitful discussions. A.L.B acknowledges support from BGU and National Science Foundation (CHE-1462075) for partial support. M.S. acknowledges financial support from the Israel Science Foundation (Grant No. 821/14). This work is supported by the European Research Council project number
	335933. 
\end{acknowledgments}
	

\begin{thebibliography}{38}
\bibitem [{Res()}]{ResNote}%
	\BibitemOpen
	\href@noop {} {}\bibinfo {note} {Another pair of coupled CPWRs was present on
		the chip but not used in the experiments.}\BibitemShut {Stop}%
\bibitem [{Pen()}]{PendulaNote}%
	\BibitemOpen
	\href@noop {} {}\bibinfo {note} {Notice that the analogy to coupled pendula
		is not complete. For example, for the pendula the symmetric mode has a lower
		energy (and frequency) than the anti-symmetric one, while the opposite is
		true for capacitively coupled resonators.}\BibitemShut {Stop}%
\end{thebibliography}
%

\end{document}


\title{Supplementary materials to "Revealing the nonlinear response of a tunneling two-level systems ensemble using coupled modes"} 
\author{Naftali Kirsh}
\affiliation{Racah Institute of Physics, the Hebrew University of Jerusalem, Jerusalem, 91904 Israel}
\author{Elisha Svetitsky}
\affiliation{Racah Institute of Physics, the Hebrew University of Jerusalem, Jerusalem, 91904 Israel}
\author{Alexander L. Burin}
\affiliation{Department of Chemistry, Tulane University, New Orleans, Louisiana 70118, USA}
\author{Moshe Schechter}
\affiliation{Department of Physics, Ben-Gurion University of the Negev, Beer Sheva 84105, Israel}
\author{Nadav Katz}
\affiliation{Racah Institute of Physics, the Hebrew University of Jerusalem, Jerusalem, 91904 Israel}
\date{\today}
\pacs{}
\maketitle

\section{Theoretical Model for frequency and loss tangent shift}
As mentioned in the text, for a small probe field one can describe the TLSs contribution to the change in the probed mode resonant frequency and loss tangent as 
\begin{equation}
	\frac{\delta\omega_{pr}}{\omega_{pr}}=\sum_{j}\frac{g_j^2}{\omega_{pr}}\frac{\langle\hat{S}_{z,j}\rangle}{\omega_{j}-\omega_{pr}+\frac{i}{T_{2j}}} ,
	\label{eq:eps_general}
\end{equation}
with coupling constant $ g_j = \frac{\Delta_{0j}}{\hbar\omega_j}\frac{p \cos(\theta_{j}) E_{N=1,j}}{\hbar} $, where $E_{N=1,j}$ is the electric field strength at the position of TLS $j$ for an average single-photon energy in the resonator. The doubled negative imaginary part of Eq.~(\ref{eq:eps_general}) determines the loss tangent due to TLSs and the real part gives the frequency shift. Here the contribution of TLSs to the dielectric constant at frequency $\omega_{pr}$ is examined, TLSs are enumerated by the letter $j$, $p_{j}$ stands for TLS $j$ dipole moment and $\theta_{j}$ is the angle between this dipole moment and the cavity field, $\Delta_{0j}$ stands for the TLS $j$ tunneling amplitude while $E_{j}=\hbar\omega_{j}$ is the energy of this TLS, the times $T_{2j}$ and $T_{1j}$ ($T_{2j} = 2T_{1j}$) describe TLS relaxation and decoherence rates which can be expressed as (weak interaction limit corresponding to low temperatures $T\sim20$ mK as in experiment) \cite{burin1995dipole}
\begin{eqnarray} 
\frac{1}{T_{1j}}=\frac{2}{T_{2j}}=A\left(\frac{\Delta_{0j}}{\hbar\omega_{j}}\right)^2 \frac{\hbar^3\omega_{j}^{3}}{k_{B}^3}\coth\left(\frac{\hbar\omega_{j}}{2k_{B}T}\right), 
\nonumber\\
A \sim 10^{8} s^{-1}K^{-3}. 
\label{eq:TLSreldec}
\end{eqnarray}
$\langle\hat{S}_{z,j}\rangle= -\Delta n_{j}$, where $\Delta n_{j}$ stands for the population difference between ground and excited states.
For a uniform electric field inside a cavity with volume V we get
\begin{eqnarray}
	\frac{\delta\omega_{pr}}{\omega_{pr}}=\sum_{j}\frac{\cos^{2}(\theta_{j})p_{j}^2}{2\hbar\epsilon\epsilon_{0}V}\left(\frac{\Delta_{0j}}{\hbar\omega_{j}}\right)^2\frac{\langle\hat{S}_{z,j}\rangle}{\omega_{j}-\omega_{pr}+\frac{i}{T_{2j}}},
	\label{eq:eps}
\end{eqnarray}
where $\epsilon$ and $\epsilon_{0}$ are environment and vacuum dielectric constants. 
If we replace $\Delta n_{j}$ with its equilibrium value  
\begin{eqnarray}
	\Delta n_{j}=\tanh\left(\frac{\hbar\omega_{j}}{2k_{B}T}\right) 
	\label{eq:TLSPopDiff}
\end{eqnarray}
Eq. (\ref{eq:eps}) will lead to the standard expressions (see Eq.~(\ref{eq:Fr1aMod1})). 

In the case of interest the presence of the second strong (pump) field affects the dielectric constant modifying population differences as 
\begin{eqnarray} 
	\Delta n_{j}=\tanh\left(\frac{\hbar\omega_{j}}{2k_{B}T}\right)
	\left[1-\frac{T_{1j}\Omega_{Rj}^{2}/T_{2j}}{(\omega_{j}-\omega_{pump})^2+\frac{1}{T_{2j}^2}\left(1+\Omega_{Rj}^2T_{1j}T_{2j}\right)}\right], 
	\label{eq:TLSPopDiffb}
\end{eqnarray}
where the TLS Rabi frequency $\Omega_{Rj}$ is given by
\begin{eqnarray}
	\Omega_{Rj}=\frac{\Delta_{0j}}{\hbar\omega_{j}}\cos(\theta_{j})\Omega_{0}, 
	\nonumber\\
	\hbar\Omega_{0}=2p_{j}F_{pump}, 
	\label{eq:TLSRabi}
\end{eqnarray}
where $F_{pump}$ is the pump field. The population difference is found by solving the standard Bloch equations \cite{Phillips}. 

The correction to the probed mode resonant frequency can be expressed through the induced change in dielectric constant due to correction to the equilibrium population number given by the second term in Eq.~(\ref{eq:TLSPopDiffb})
\begin{widetext}
	\begin{eqnarray}
		\frac{\delta\omega_{pr}}{\omega_{pr}\tanh\left(\frac{\hbar\omega_{pr}}{2k_{B}T}\right)}=
		\nonumber\\
		=-\frac{P_{0}}{2\epsilon\epsilon_{0}}\int_{0}^{1}dy\int_{0}^{\infty} d\omega \int_{0}^{1} \frac{d\xi}{\xi\sqrt{1-\xi^2}}\frac{\xi^2y^2 p^2}{\omega_{pr}-\omega-\frac{i}{T_{2}}} 
		\frac{\xi^2y^2\Omega_{0}^2/2}{(\omega-\omega_{pump})^2+\frac{1}{T_{2}^2}\left(1+\xi^2y^2\Omega_{0}^2T_{2}^2/2\right)}
		\nonumber\\
		\xi=\frac{\Delta_{0}}{\hbar\omega}, ~ y=\cos(\theta).
		\label{eq:Fr1}
	\end{eqnarray}
\end{widetext}
The integration over TLS frequencies $\omega$ can be performed analytically extending it to the negative infinity. Then we get 
\begin{widetext}
	\begin{eqnarray}
		\frac{\delta\omega_{pr}}{\omega_{pr}\tanh\left(\frac{\hbar\omega_{pr}}{2k_{B}T}\right)}=
		\nonumber\\
		=-\frac{P_{0}\pi}{4\epsilon\epsilon_{0}}\int_{0}^{1}dy\int_{0}^{1} \frac{d\xi}{\xi\sqrt{1-\xi^2}}\frac{\xi^4y^4 p^2\Omega_{0}^2T_{2}}{\omega_{pr}-\omega_{pump}-\frac{i}{T_{2}}\left[1+\sqrt{1+\xi^2y^2\Omega_{0}^2T_{2}^2/2}\right]} \frac{1}{\sqrt{1+\xi^2y^2\Omega_{0}^2T_{2}^2/2}}.
		\label{eq:Fr1a}
	\end{eqnarray}
\end{widetext}

In the case of zero frequency difference $\left(\omega_{pump}=\omega_{pr}\right)$ this equation should approach the standard model behavior. To show that this is indeed true one can rewrite this expression as 
\begin{widetext}
	\begin{eqnarray}
		\frac{\delta\omega_{pr}}{\omega_{pr}\tanh\left(\frac{\hbar\omega_{pr}}{2k_{B}T}\right)}
		=-i\frac{P_{0}\pi}{2\epsilon\epsilon_{0}}\int_{0}^{1}dy\int_{0}^{1} \frac{p^2y^2 \xi^2d\xi}{\xi\sqrt{1-\xi^2}}
		\nonumber\\
		\left[1+\left(\frac{\xi^2y^2 p^2\Omega_{0}^2T_{2}/2}{\frac{1}{T_{2}}\left[1+\sqrt{1+\xi^2y^2\Omega_{0}^2T_{2}^2/2}\right]} \frac{1}{\sqrt{1+\xi^2y^2\Omega_{0}^2T_{2}^2/2}}-1\right)\right]=
		\nonumber\\
		=-\frac{i}{2}\frac{P_{0}<p^2>\pi}{3\epsilon\epsilon_{0}}+\frac{i}{2}\frac{P_{0}\pi}{\epsilon\epsilon_{0}}\int_{0}^{1}dy\int_{0}^{1} \frac{p^2y^2 \xi^2 d\xi}{\xi\sqrt{1-\xi^2}}\frac{1}{\sqrt{1+\xi^2y^2\Omega_{0}^2T_{2}^2/2}}
		\label{eq:Fr1aMod1}
	\end{eqnarray}
\end{widetext}
The unity term in brackets yields $-\frac{i}{2}\frac{P_{0}<p^2>\pi}{3\epsilon\epsilon_{0}}$ which is one half of the negative loss tangent corresponding to the linear response theory. It cancels out the main contribution corresponding to the first term in Eq. (\ref{eq:TLSPopDiffb}) in the large field limit. The second term represents the exact expression for $1/2$ of the non-linear loss tangent within the standard TLS model. 

\begin{figure}[h!]
	\subfloat[]{
	\centering
	\includegraphics[trim={1.2cm 6.3cm 1.5cm 9cm},clip,width=0.49\linewidth]{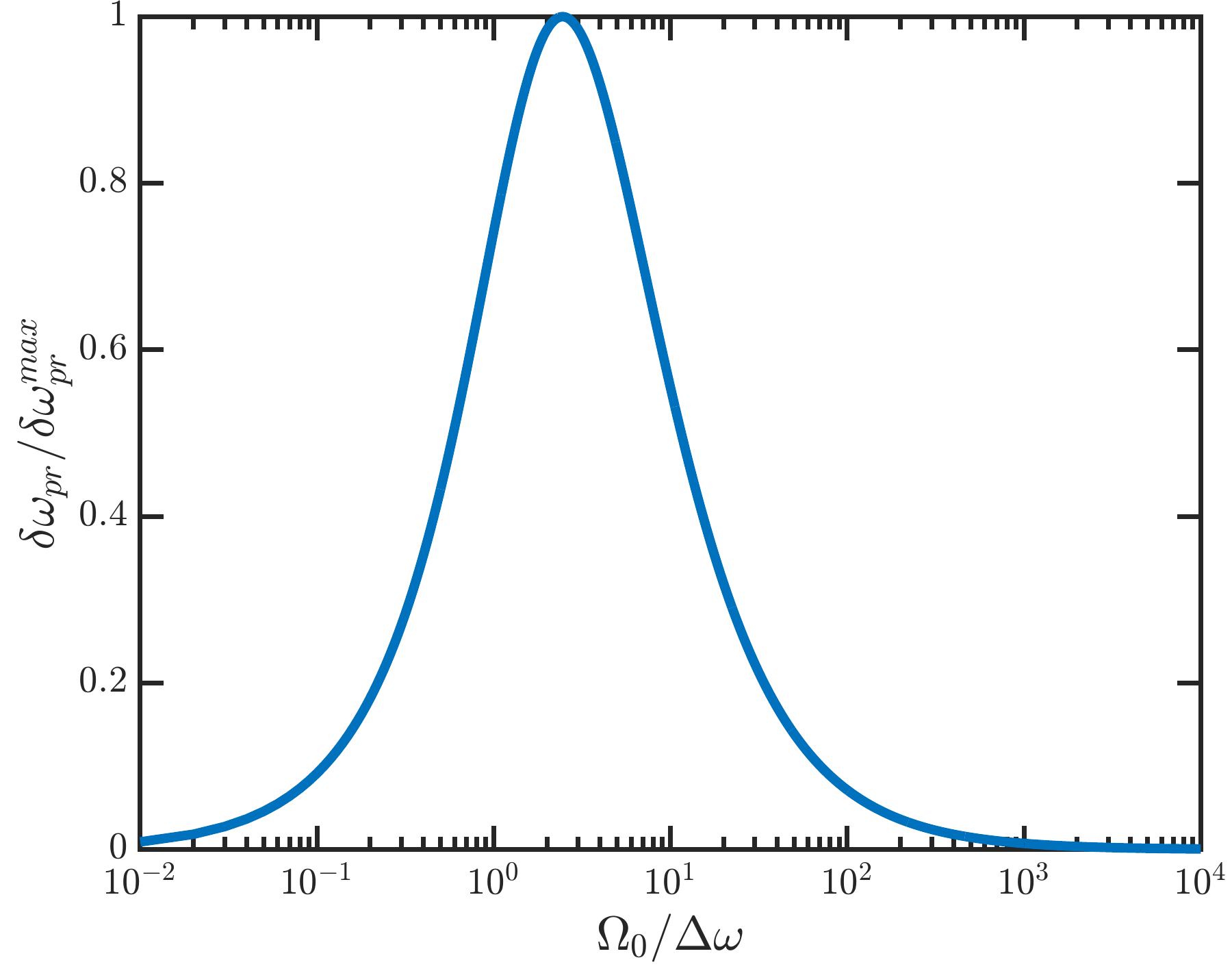}
	\label{fig:Shift}
	}
	\subfloat[]{
	\centering
	\includegraphics[trim={1.2cm 6.3cm 1.5cm 9cm},clip,width=0.49\linewidth]{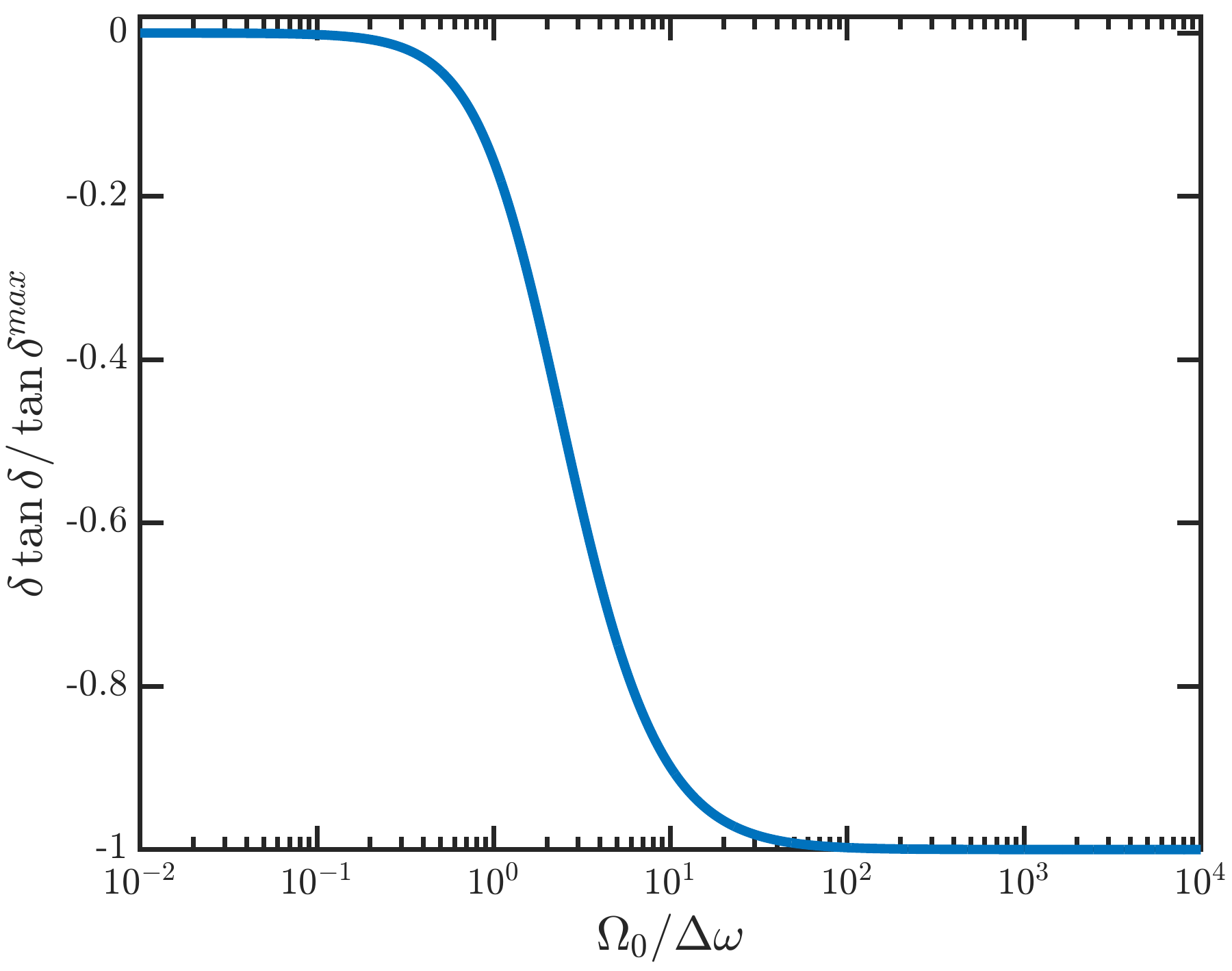}
	\label{fig:ShiftLT}
	}
	\caption{Theoretical calculation of (a) the frequency shift and (b) the loss tangent, as a function of external field (for positive $ \Delta\omega $).}
\end{figure}

In the limit of interest of large non-linearity $\Omega T_{1,2} \gg 1$ where the correction becomes really large one can simplify Eq. (\ref{eq:Fr1a}) as 
\begin{widetext}
	\begin{eqnarray}
		\frac{\delta\omega_{pr}}{\omega_{pr}\tanh\left(\frac{\hbar\omega_{pr}}{2k_{B}T}\right)}=
		\nonumber\\
		=\frac{\sqrt{2}P_{0}\pi}{4\epsilon\epsilon_{0}}\int_{0}^{1}dy\int_{0}^{1} \frac{d\xi}{\sqrt{1-\xi^2}}\frac{\xi^2y^3 p^2\Omega_{0}}{\Delta\omega+\frac{i\xi y\Omega_{0}}{\sqrt{2}}}, ~ \Delta\omega = \omega_{pump}-\omega_{pr}.
		\label{eq:Fr2}
	\end{eqnarray}
\end{widetext}
The real part of the correction to the frequency can be evaluated exactly as 
\begin{widetext}
	\begin{eqnarray}
		\frac{\delta\omega_{pr}}{\omega_{pr}\tanh\left(\frac{\hbar\omega_{pr}}{2k_{B}T}\right)}=
		\frac{\sqrt{2}P_{0}p^2\pi^2}{8\epsilon\epsilon_{0}}\frac{\Delta\omega}{\Omega_{0}}\frac{\sqrt{1+\frac{\Omega_{0}^2}{2\Delta\omega^2}}-1}{\sqrt{1+\frac{\Omega_{0}^2}{2\Delta\omega^2}}+1}.
		\label{eq:Fr3}
	\end{eqnarray}
\end{widetext}
This result expresses the field dependence of the frequency shift in terms of the maximum 
Rabi frequency $\Omega_{0}=2pF_{pump}/\hbar$. The result is sensitive to the distribution of dipole moments $p$ absolute value. 
Assuming the dipole moment magnitude to be approximately constant (see e.g. Ref.~\cite{MartinisTLS}, but see Ref.~\cite{twoTypesOsborn2} and the discussion below) one can predict that the frequency shift increases with the field at small fields $F_{pump} \ll \hbar\Delta\omega/p$ where one can expand the numerator with respect to the small ratio $\Omega_{0}/\Delta\omega$  as 
\begin{equation}
	\frac{\delta\omega_{pr}^{low}}{\omega_{pr}} = \frac{\sqrt{2}P_{0}p^2\pi^2}{64\epsilon\epsilon_{0}}\frac{\Omega_{0}}{\Delta\omega},
	\label{eq:lowFieldShift}
\end{equation}
reaches the maximum $|\delta\omega_{pr}^{max}/\omega_{pr}|=\frac{\pi^2}{24\sqrt{3}}\frac{P_{0}p^2}{\epsilon\epsilon_{0}}$ at $\Omega_{0}^{max}=\sqrt{6}|\Delta\omega|$ and then decreases as
\begin{equation}
	\frac{\delta\omega_{pr}^{high}}{\omega_{pr}} = \frac{\sqrt{2}P_{0}p^2\pi^2}{8\epsilon\epsilon_{0}}\frac{\Delta\omega}{\Omega_{0}}
\end{equation}
with increasing the field at large fields $F_{pump} \gg \hbar\Delta\omega/p$ (where we assumed $\tanh\left(\frac{\hbar\omega_{pr}}{2k_{B}T}\right)\approx 1$ as in our experiments). The theory predictions are illustrated in Fig. \ref{fig:Shift}. The maximum frequency shift can be conveniently expressed in terms of the weak-field loss tangent $ \tan \delta_0 = \frac{\pi P_0 p^2}{3\epsilon\epsilon_{0}} $  as 
\begin{eqnarray}
	| \frac{\delta\omega_{max}}{\omega_{pr}\tan\delta_0} | =\frac{\pi}{8\sqrt{3}}\approx 0.227.
	\label{eq:MaxShift}
\end{eqnarray}
In our experiment we obtain $ \frac{\delta\omega_{max}}{\omega_{pr}\tan\delta_0} \approx 0.23 $ when the probe is at 5689 MHz which agrees with theory, but when the probe is at 5626 MHz we get      
$ \frac{\delta\omega_{max}}{\omega_{pr}\tan\delta_0} \approx 0.16 $, which deviates from the uniform-field theory (here we denote $1/Q_i = \tan\delta$ even for our case of a non-uniform electric field, see below). We note that in a Monte-Carlo simulation which takes into account the finite number of TLSs and the non-uniform electric field (see below) we see a variation in the value of $|\frac{\delta\omega_{max}}{\omega_{pr}\tan\delta_0}|$ between different realizations and we can indeed see values close to $0.16$. In addition, one should notice that the present estimate relies on the assumption of the constant absolute value of the TLS dipole moment, a broader distribution of dipole moments will definitely lead to the reduction of this maximum since it will mix up the contribution from the maximum Rabi frequency with smaller contributions of other Rabi frequencies.

If the frequency detuning $\Delta \omega$ approaches zero (corresponding to one-tone experiments) the fraction $\frac{\sqrt{1+\frac{\Omega_{0}^2}{2\Delta\omega^2}}-1}{\sqrt{1+\frac{\Omega_{0}^2}{2\Delta\omega^2}}+1}$ tends to unity and the frequency shift approaches zero as $\Delta \omega$ . 

The imaginary part of the relative frequency shift representing $1/2$ of the loss tangent correction can be also calculated using Eq. (\ref{eq:Fr2}). The associated correction to the loss tangent can be expressed as 
\begin{widetext}
	\begin{eqnarray}
		\frac{\delta\tan(\delta)}{\tanh\left(\frac{\hbar\omega_{pr}}{2k_{B}T}\right)}\bigg/\frac{\pi P_{0}p^2}{3\epsilon\epsilon_{0}} =
		\nonumber\\
		=
		-\left[1+\left(\frac{\Delta\omega}{\Omega_0}\right)^2\left\{6+3\sqrt{1+2\left(\frac{\Delta\omega}{\Omega_0}\right)^2}
		\ln\left(1+\left(\frac{\Omega_0}{\Delta\omega}\right)^2\left[1-\sqrt{1+2\left(\frac{\Delta\omega}{\Omega_0}\right)^2}\right]\right)\right\}\right]
		\label{eq:LossTan}
	\end{eqnarray}
\end{widetext}
The expected dependence of the loss tangent on the external field amplitude is shown in Fig. \ref{fig:ShiftLT}. 

If the TLSs' hosting material doesn't fill all the medium Eq.~(\ref{eq:Fr3}) and~(\ref{eq:LossTan}) should be multiplied by the appropriate filling factor \cite{GaoThesis}. In addition, for a mode with a non-uniform electric field distribution the internal loss $ 1/Q_i $ is in general different from the bulk loss tangent $ \tan \delta $. Nevertheless, since, as we show below, there is a nearly identical power dependence of $ 1/Q_i $ and $ \tan \delta $ in the regime relevant for our experiments we use the notation $ 1/Q_i \equiv \tan \delta  $ throughout this appendix.

\section{Calculating the steady-state photon number in the resonator}
The steady state photon number in a resonator with resonance frequency $ \omega_0 $ driven at power $ P $ with a pump tone of frequency $ \omega_p $  is given by \cite{Optomechanics}
\begin{equation}
	\langle N \rangle = \frac{\kappa_{ex}}{\left(\kappa/2\right)^2+\Delta^2}\frac{P}{\hbar\omega_p}.
\end{equation}
Using the definitions of external and total decay rates $ \kappa_{ex} \equiv \frac{\omega_0}{Q_c} $, $ \kappa \equiv \frac{\omega_0}{Q_l}  $ and the detuning between the drive and the resonator $ \Delta \equiv \omega_p - \omega_0 $ we obtain 
 \begin{equation}
	 \langle N \rangle = 4\frac{Q_l}{\omega_0}\frac{Q_l}{Q_c}\frac{1}{1+\left(2Q_l\Delta/\omega_0\right)^2}\frac{P}{\hbar\omega_p}.
 \end{equation} 
$ Q_l,Q_c $ and $ \omega_0 $ were extracted from the fits. For the pump-and-probe experiments $ \omega_p $ was constant while the pumped-mode resonance frequency $ \omega_0=\omega_0(P) $ is power-dependent, hence $ \Delta $ was calculated using $ \omega_0(P) $ which was obtained from the probe-only experiments. We note that this calculation ignores impedance mismatches on the device and at the input chain of the fridge \cite{QcNote} and therefore the calculated $ \langle N \rangle $ cannot be considered completely accurate. A knowledge of the exact number of photons is possible only by coupling the resonator to a nonlinear quantum system \cite{schusterResolvingN}. Nevertheless, indication that our calculated $ \langle N \rangle $ is close to the real value can be found by (a) the agreement of the extracted nonlinear Kerr coefficients with order-of-magnitude estimations (see below) and (b) the agreement between the pumping number of photons for a maximal frequency shift and theory as detailed now. According to our theoretical derivation, the maximal frequency shift occurs at $ \Omega_0^{max} = \sqrt{6}\Delta\omega $, where $ \Omega_0 = \Omega_0^{N=1}\sqrt{N} $ with $ \Omega_0^{N=1} $ the single photon Rabi frequency. 
$\Omega_0^{N=1}$ is calculated by fitting the two-tone frequency shift (see next section) which give $\Omega_0^{N=1}\approx 2\pi\times 79$ kHz. 
Substituting $ \Delta\omega = 2\pi\times 63$ MHz 
we obtain that the maximal frequency shift should occur at
\begin{equation}
	\langle N \rangle^{max} = 6\left(\frac{\Delta\omega}{\Omega_0^{N=1}}\right)^2 \approx 3.8\times 10^6,
\end{equation} 
which is in the order of the value we measured (see Fig.~3a in the main manuscript, but notice that the actual maximal shift is measured by first subtracting the nonlinear kinetic inductance shifts, see next section). 
In any case we stress that the exact value of $ \langle N \rangle $ does not effect our qualitative results.   

\section{Estimating the average single photon Rabi frequency of a dominant TLS}
Using  Eq.~(\ref{eq:Fr3}) and knowing that the maximal shift occurs at $\Omega_{0}^{max}=\sqrt{6}|\Delta\omega|$ we obtain
\begin{equation}
	\frac{\delta\omega_{pr}}{\delta\omega_{pr}^{max}} = 3\sqrt{6}\frac{\Delta\omega}{\Omega_{0}}\frac{\sqrt{1+\frac{\Omega_{0}^2}{2\Delta\omega^2}}-1}{\sqrt{1+\frac{\Omega_{0}^2}{2\Delta\omega^2}}+1}.
	\label{eq:norm_shift}
\end{equation}    
Since $ \Omega_0 = \Omega_0^{N=1}\sqrt{N} $ with $ \Omega_0^{N=1} $ the single photon Rabi frequency and because we know the detuning $ \Delta\omega $ we can fit our two-tone frequency measurements vs. $ \langle N \rangle  $ to Eq.~(\ref{eq:norm_shift}) and extract $ \Omega_0^{N=1} $. In order to exclude the nonlinear kinetic inductance shifts we first subtract them using the fitted nonlinear Kerr coefficients (notice that reasonable results are obtained only when we use the value of $ K^\prime $ fitted for the 5689 MHz resonance, supporting the assumption that the value of $ K^\prime $ extracted from fitting the 5626 MHz resonance is not the correct one (see main manuscript)). This yields $\Omega_0^{N=1}/2\pi \approx 81$ kHz when probing at 5626 MHz and $\Omega_0^{N=1}/2\pi \approx 78$ kHz when the probe is at at 5689 MHz. These values which are much larger than the ones expected for a uniform electric field ($\Omega_0^{N=1}/2\pi\approx 5.2$ kHz) confirm the predictions of previous studies \cite{Xmon,MartinisSimulations} and of our Monte-Carlo simulations (see below) that TLSs at regions of strong fields dominate the nonlinear behavior.           
\section{Monte-Carlo simulations}
In order to check the effects of a finite number of TLSs and non-uniform electric fields we have conducted numerical Monte-Carlo simulations (a similar approach was used in Ref.~\cite{Xmon}). First, we calculate the electric field distribution of a CPWR using the potential matrix method \cite{PotentialMatrix}. The expected number of TLSs $ N_{TLS}=P_0 V \hbar 2\pi B $ is calculated using the density of states $ P_0 = 10^{45}\,J^{-1}m^{-3}$ \cite{Berret1988}, a bandwidth of $B=2$ GHz and our geometrical dimensions, where we assumed that TLSs are located at a distance of 2.5 nm from metal-air/metal-dielectric and substrate-air interfaces \cite{GaoSurface,MartinisSimulations}. For the substrate-air interface we assumed TLS to be inside the substrate and for the metal-air we assumed that they are inside a dielectric layer with $\epsilon = 10$. In order to make the calculation computationally reasonable we first calculated the field in absence of the additional thin layer and then used the appropriate electric field boundary conditions to calculate the fields the TLSs experience \cite{MartinisSimulations}. For each numerical realization TLSs are randomly placed inside these layers and their frequency $\omega_j$ coupling strength $\left(\frac{\Delta_{0j}}{\hbar\omega_j}\right)^2$ and relative angle to the field $\theta_j$ are acquired using the standard distributions \cite{Phillips,hunklinger1988tunneling}. Then, the electric field is calculated yielding the maximal Rabi frequency $\Omega_0 = \frac{2pE_{N=1,j}}{\hbar}\sqrt{N}$ and the coupling constant $ g_j = \frac{\Delta_{0j}}{\hbar\omega_j}\frac{p \cos(\theta_{j}) E_{N=1,j}}{\hbar} $, where $E_{N=1,j}$ is the calculated electric field at the position of TLS $j$ for an average single-photon energy in the resonator. We assumed that the dipole moment magnitude is a constant with the value $P=2.8$ D \cite{twoTypesOsborn2}. 
The frequency shift and half of the loss tangent are calculated by performing the sum Eq.~(\ref{eq:eps_general}) and taking the real and imaginary parts respectively. The numerical codes are available for download in our group's web site \cite{GroupSite}. We stress that these simulations are not intended for estimating the absolute values of the loss and frequency shifts, since these depend on some values which are not precisely known such as TLS density of states $P_0$ and can also depend on the numerical grid size. Adaptive meshing which is required for exactly calculating the fields at nanometric distances in a micron-size geometry is outside the scope and need of this Letter.  
\begin{figure}[h!]
	\subfloat[]{
		\centering
		\includegraphics[trim={1cm 6cm 1cm 9cm},clip,scale=0.45]{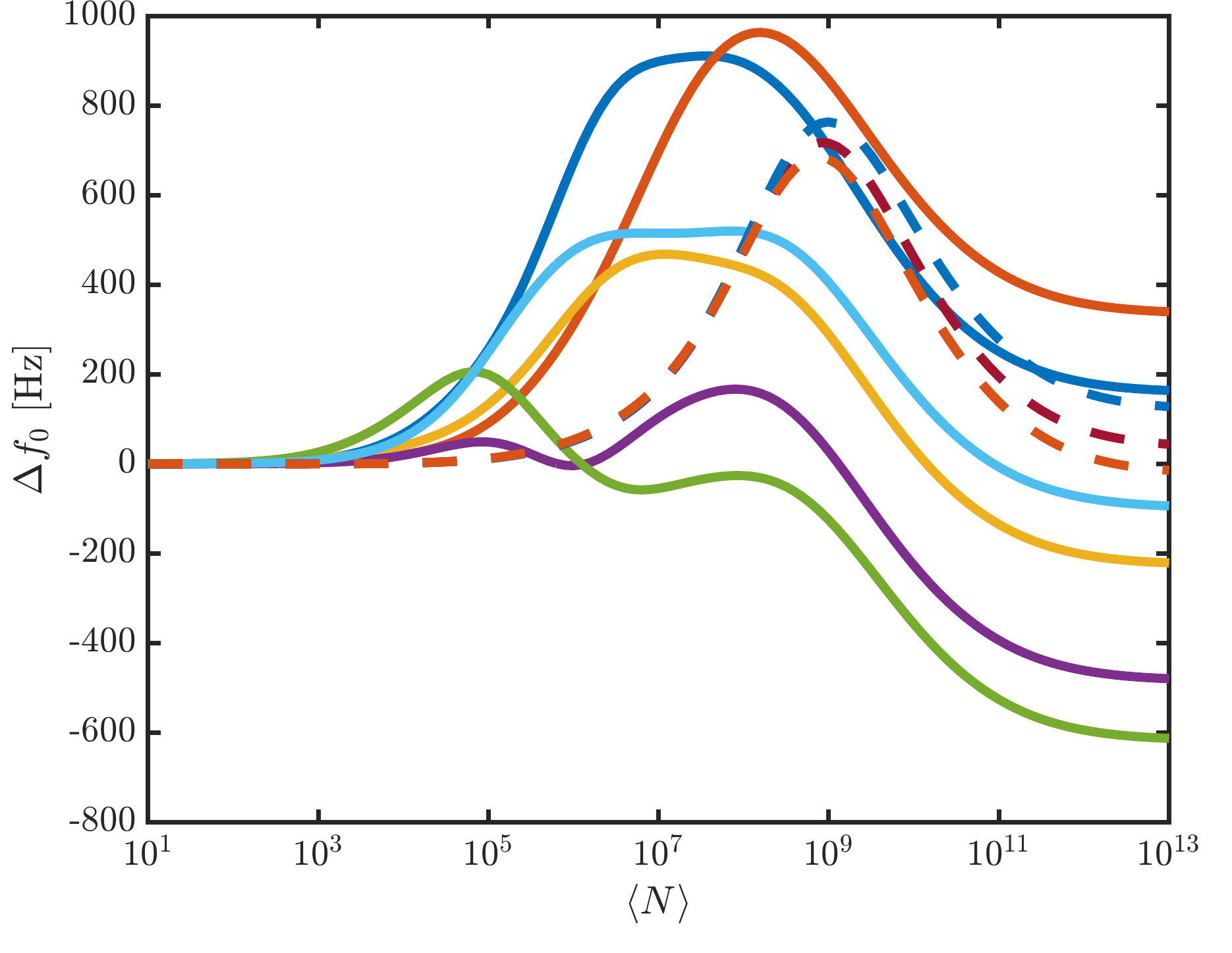}
		\label{Simul2ToneShift}
	}
	\subfloat[]{
		\centering
		\includegraphics[trim={1cm 6cm 1cm 9cm},clip,scale=0.45]{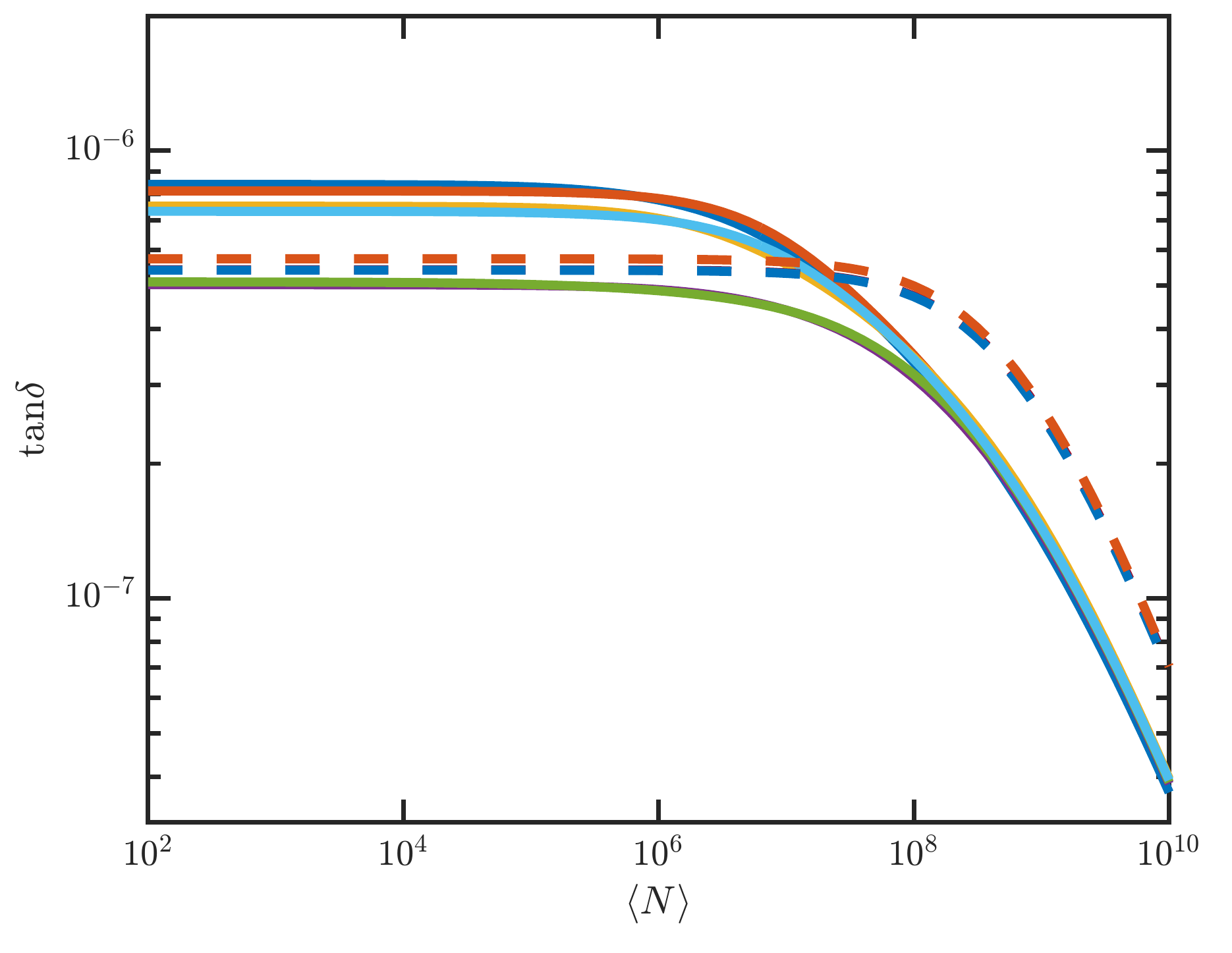}
	}
	\caption{Simulated (a) frequency shifts and (b) loss tangents as a function of pump photons (for a positive $ \Delta\omega/2\pi = 60 $ MHz). Each line corresponds to a different numerical realization. Full (Dashed) lines show simulations with non-uniform (uniform) electric fields.}
	\label{fig:Simul2ToneResults}
\end{figure}
In Fig ~\ref{fig:Simul2ToneResults} we show the simulated shifts of frequency and loss tangent as a function of pump photons assuming a uniform electric field (dashed lines) and when the non-uniformity of the field is taken into account (full lines). These are simulations of pump-and-probe experiments with $ \Delta\omega / 2\pi = 60$ MHz. In Fig.~\ref{fig:Simul2ToneFit} we show fits of the simulation results to the theoretical model Eq.~(\ref{eq:Fr3}) for both uniform and non-uniform electric field.   
\begin{figure}[h!]
	\subfloat[]{
		\centering
		\includegraphics[scale=0.35]{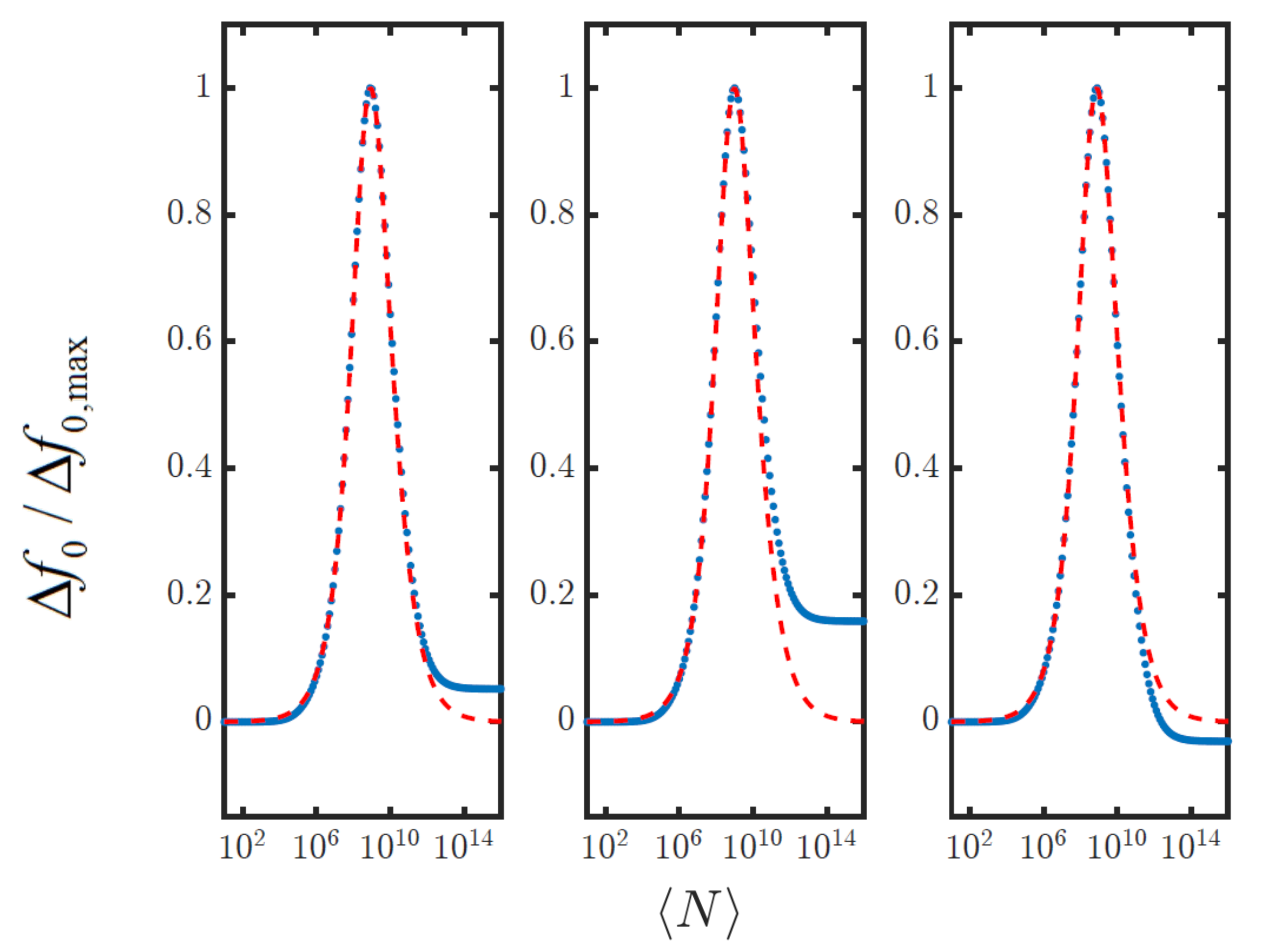}
		\label{fig:Simul2ToneFitU}
	}
	\subfloat[]{
		\centering
		\includegraphics[scale=0.35]{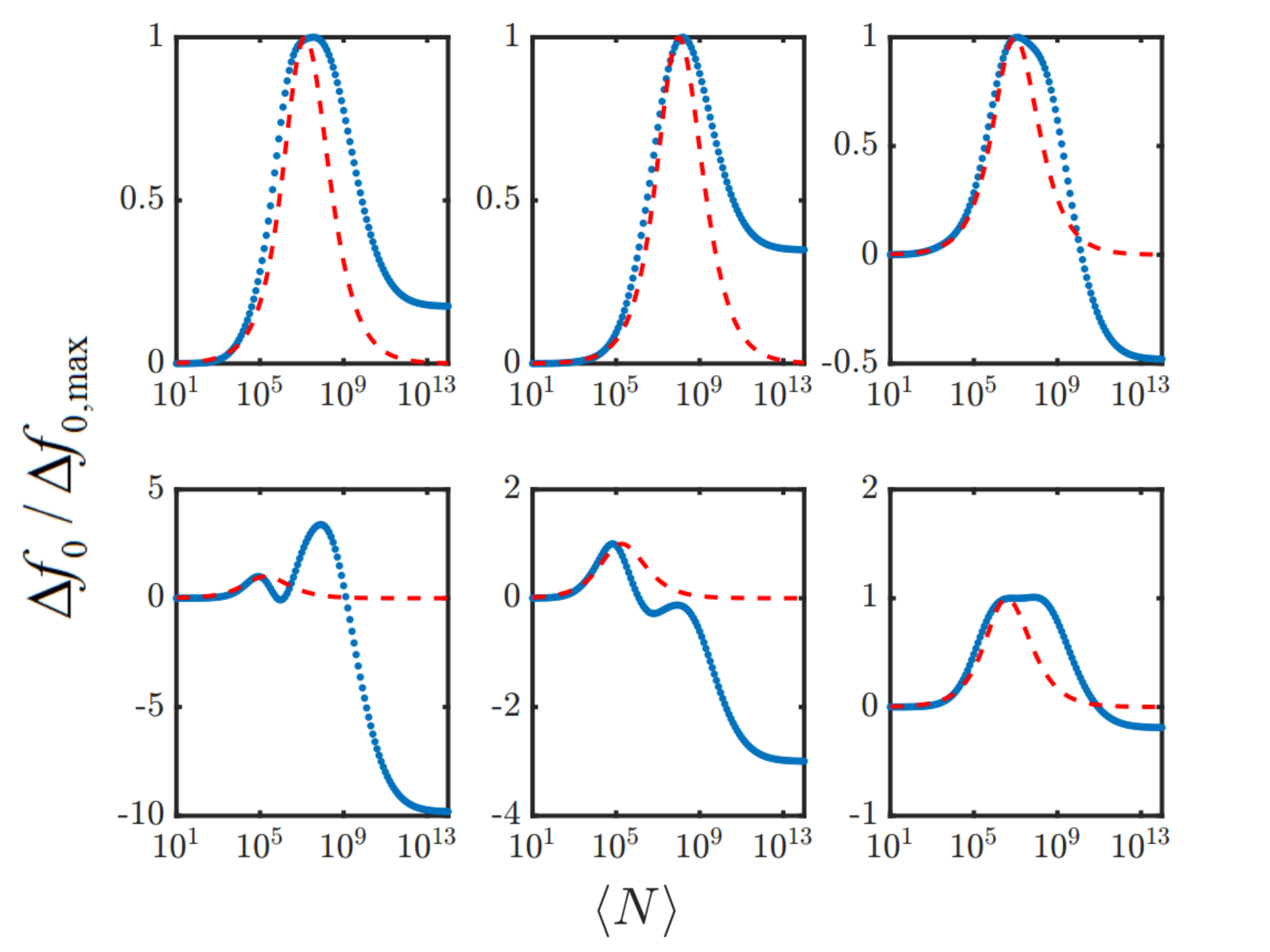}
		\label{fig:Simul2ToneFitNonU}
	}
	\caption{Fitting the simulated frequency shifts to the theoretical model for (a) uniform and (b) non-uniform electric fields. Each sub-figure corresponds to a different numerical realization (same as those of Fig.~\ref{fig:Simul2ToneResults}). Blue dots are the simulation results, red dashed lines are fits to Eq.~(\ref{eq:Fr3}).}
	\label{fig:Simul2ToneFit}
\end{figure} 
As can be seen in Fig.~\ref{fig:Simul2ToneFitU} and Fig.~\ref{Simul2ToneShift}, for a uniform electric field the theory fits the simulations very well except from some variance between the different realizations and a non-vanishing shift which remains for high pump powers. Both effects are explained by the finite number of TLS which in addition to the variance results in a finite asymmetry even at the absence of pumping. Due to the asymmetry for each realization there is a different initial frequency shift which translates to a different $\Delta f_0$ at high powers. In contrast to the uniform field case, if the non-uniformity of the field is taken into account, the theory fits the simulations qualitatively but can have different amounts of quantitative deviations as shown in Fig.~\ref{fig:Simul2ToneFitNonU}. In rare cases (not shown) the qualitative picture breaks as well. In addition, there are large differences between different realizations and the behavior is very irregular. The reason for these deviations is that the shifts are dominated by a few TLSs which are located in regions of strong electric fields such a corners \cite{Xmon,MartinisSimulations}. These dominating TLSs cause also that the fitted maximal single-photon Rabi frequency  $\Omega_0^{N=1}$ will be much larger than the value expected for a uniform field. For example, for the three realizations shown in the top row of Fig.~\ref{fig:Simul2ToneFitNonU} which are reasonably fitted the extracted single-photon Rabi frequencies are  $\Omega_0^{N=1}/2\pi=39$, $15$  and $49$ kHz , while for a uniform field we expect $\Omega_0^{N=1}/2\pi\approx 5.2$ kHz. The uniform-field simulations indeed give $\Omega_0^{N=1}/2\pi=5$, $4.7$ and $5.4$ kHz which are close to the theoretical value. Large values of $\Omega_0^{N=1}$ were also fitted from our experiments' data (see main manuscript) confirming the assumptions that strongly coupled TLSs dominate the shifts. In addition, the differences between the shifts of both resonances and small deviation from the theoretical curves (see Fig.~3a of the main manuscript) are explained by the effect of small dominating TLSs as confirmed by the simulations.  

\begin{figure}[h!]
	\subfloat[]{
		\centering
		\includegraphics[trim={1cm 6cm 1cm 7cm},clip,scale=0.45]{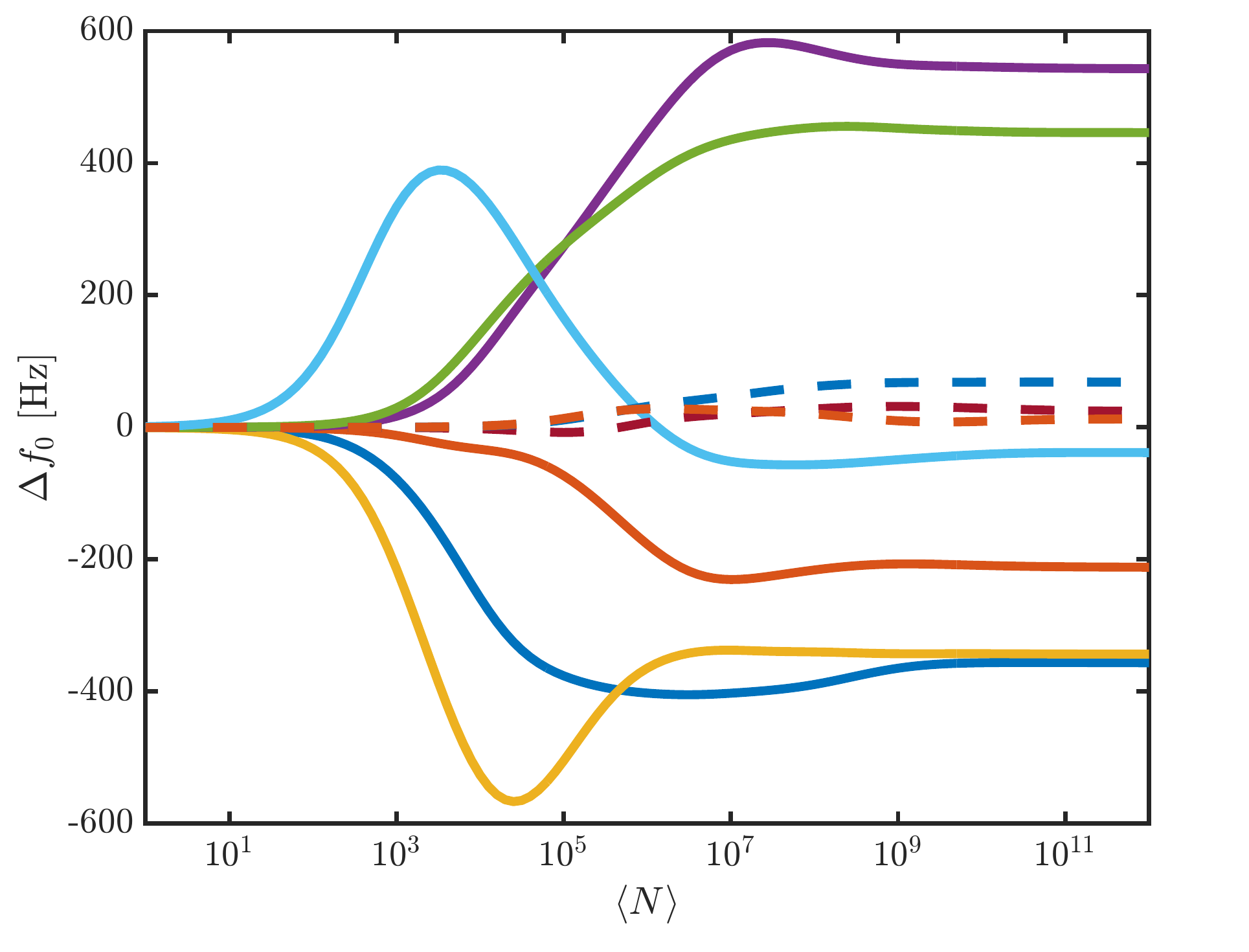}
		\label{fig:SimulOneToneShift}
	}
	\subfloat[]{
		\centering
		\includegraphics[trim={1cm 6cm 1cm 7cm},clip,scale=0.45]{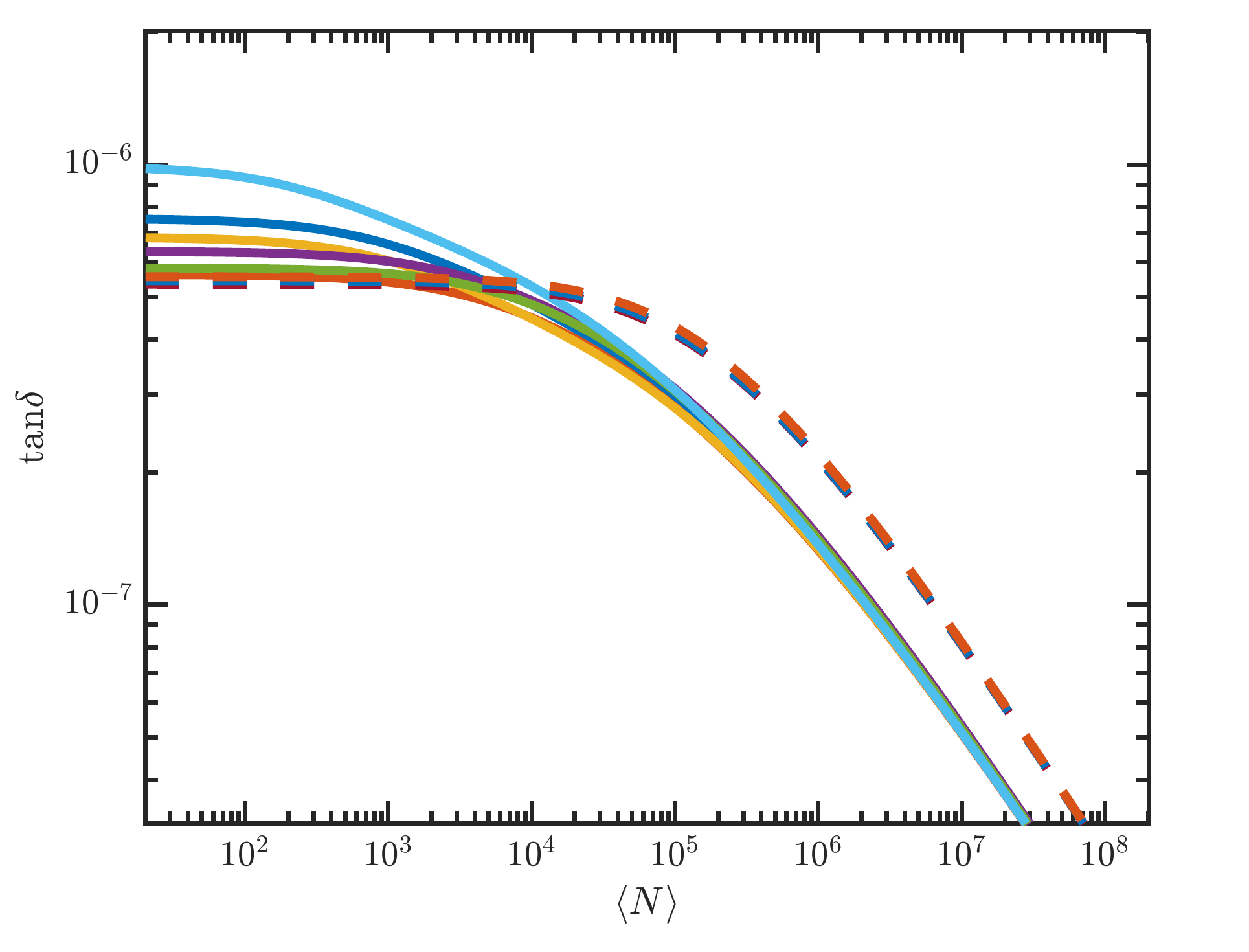}
		\label{fig:SimulOneToneLoss}
	}
	\caption{Simulated (a) frequency shifts and (b) loss tangents of probe-only experiments. Full (Dashed) lines show simulations with non-uniform (uniform) electric fields.}
	\label{fig:SimulOneToneResults}
\end{figure}

In addition to the simulations of two-tone experiments we also checked the effect of finite number of TLSs and non-uniform electric fields on one-tone probe-only measurements. The simulations results are shown in Fig.~\ref{fig:SimulOneToneResults}. As can be seen in Fig.~\ref{fig:SimulOneToneLoss} the non-uniform fields reduce the critical photon number in which saturation starts and soften the loss curves \cite{MartinisSimulations} but does not significantly effect the functional behavior at high-powers \cite{MartinisSimulations,KhalilTLSloss}. In addition, as mentioned above, due to the finite number of TLSs there is some asymmetry in TLSs spectral distribution around the probed resonance, resulting in frequency shifts in one-tone experiments as shown in Fig.~\ref{fig:SimulOneToneShift}. This effect is more pronounced for a non-uniform electric field in which a few TLSs in regions of strong fields have a large effect. Notice that in contrast to the two-tone case (Fig.~\ref{Simul2ToneShift}) the direction of the shift is random. This finite asymmetry can explain the low-power frequency shifts in one-tone experiments shown in Fig.~2a of the main manuscript. We notice that low-power frequency shifts in random directions were observed in another experiment of uncoupled resonators on Sapphire (unpublished).
\section{A Kerr nonlinearity model for coupled resonators}
In this section we extend the model of Yurke and Buks \cite{YurkeBuks} to the case of coupled resonators. Modeling the nonlinear inductance as a Kerr-type one we write the Hamiltonian of the coupled resonators as
\begin{equation}
H = \hbar\omega_0 a_1^{\dagger}a_1+\hbar\omega_0 a_2^{\dagger}a_2+\hbar g
\left(a_1^{\dagger}a_2+a_2^{\dagger}a_1\right)+\frac{\hbar}{2}K(a_1^{\dagger}a_1^{\dagger}a_1
a_1+a_2^{\dagger}a_2^{\dagger}a_2
a_2).
\end{equation}
The first two terms are the linear parts of both resonators (assumed to have
the same bare frequency $\omega_0$), the third term is the coupling term with
coupling $g$ and the last term contains the Kerr nonlinearity part for both
resonators, with a Kerr coefficient $K$ (again, assumed to be identical for both resonators). Defining the normal modes as
\begin{equation}
a_{\pm} \equiv \frac{a_1 \pm a_2}{\sqrt{2}}\Rightarrow a_{1,2} \equiv
\frac{a_+ \pm a_{-}}{\sqrt{2}},
\end{equation}
which obey the canonical commutation relations
\begin{equation}
\label{eq:commRel}
\left[a_i^{\dagger},a_j\right] = \delta_{ij}\,\,\left(i,j=+,-\right),
\end{equation}
we obtain the standard transformation to the normal modes
\begin{equation}
H_{lin} \equiv \hbar\omega_0 a_1^{\dagger}a_1+\hbar\omega_0 a_2^{\dagger}a_2+\hbar g
\left(a_1^{\dagger}a_2+a_2^{\dagger}a_1\right) =
\hbar(\omega_0+g)a_{+}^{\dagger}a_{+}+\hbar(\omega_0-g)a_{-}^{\dagger}a_{-}.
\end{equation}
In order to calculate the nonlinear term it is useful to notice that in the sum
\begin{eqnarray}
	a_1^{\dagger}a_1^{\dagger}a_1 a_1+a_2^{\dagger}a_2^{\dagger}a_2 a_2\\
	=\frac{1}{4}\left(a_+^\dagger+a_-^\dagger\right)\left(a_+^\dagger+a_-^\dagger\right)\left(a_{+}+a_-\right)\left(a_{+}+a_-\right) \nonumber\\
	+\frac{1}{4}\left(a_+^\dagger-a_-^\dagger\right)\left(a_+^\dagger-a_-^\dagger\right)\left(a_{+}-a_-\right)\left(a_{+}-a_-\right) \nonumber
\end{eqnarray}
only terms which are positive in the last line (i.e. which involve an even
number of minus signs) will survive, getting a factor of 2, the others will
be canceled. The result of the calculation is 
\begin{eqnarray}
	H_{nonlin} \equiv \frac{\hbar}{2}K(a_1^{\dagger}a_1^{\dagger}a_1
	a_1+a_2^{\dagger}a_2^{\dagger}a_2
	a_2)\\ 
	= \frac{\hbar}{4}K\left(a_+^{\dagger}a_+^{\dagger}a_+
	a_+ + a_-^{\dagger}a_-^{\dagger}a_-
	a_- +4a_+^{\dagger}a_+a_-^{\dagger}
	a_- + a_+^{\dagger}a_+^{\dagger}a_-
	a_- + a_-^{\dagger}a_-^{\dagger}a_+
	a_+ \right). \nonumber
\end{eqnarray}
The first two terms in the parentheses are the Kerr terms of the normal modes,
the third term is an energetic price for having photons in more than one
mode (notice that the two modes occupy the same spatial volume)
and the last two terms are two-photon exchange terms. Since from the commutation relations Eq.~(\ref{eq:commRel})
\begin{equation}
a_i^{\dagger}a_i^{\dagger}a_i a_i = a_i^{\dagger}a_i a_i^{\dagger} a_i -
a_i^{\dagger}a_i \equiv N_i(N_i-1),
\end{equation}
where $N_i$ is the number operator of mode i, we can finally write the
Hamiltonian as
\begin{eqnarray}
	H =
	\hbar(\omega_0+g)N_{+}+\hbar(\omega_0-g)N_{-}+\frac{\hbar}{4}K\left(N_+-1\right)N_{+}+\frac{\hbar}{4}K\left(N_{-}-1\right)N_{-}\\
	+\hbar K N_{-}N_{+}+\frac{\hbar}{4}K\left(a_+^{\dagger}a_+^{\dagger}a_-
	a_- + a_-^{\dagger}a_-^{\dagger}a_+
	a_+ \right) \nonumber 
\end{eqnarray}
Writing the Heisenberg equations of motion for the normal modes (neglecting non-resonant terms, assuming $K\langle N \rangle \ll 2g$ such as in our case)
\begin{equation}
	\frac{da_{\pm}}{dt}=-\frac{i}{\hbar}\left[a_{\pm},H\right]=-i\left(\left[\omega_0 \pm g\right]+\frac{K}{2}N_{\pm}+KN_{\mp}\right)a_{\pm}
\end{equation}
it can be seen that the cross-Kerr frequency shift is twice the self-Kerr one. 
\section{Fitting one-tone loss to various TLS models}
The standard TLS model predicts a square-root dependence of the loss tangent on the resonator internal energy \cite{standardLoss}, i.e. $\tan\,\delta \propto \left(1+\frac{N}{N_C}\right)^{-0.5}$ where $N_C$ is the critical photon number for saturation. While matching results of measurements on lossy dielectrics, this model didn't fit many measurements of resonators with higher internal quality factors where a phenomenological  power-law: $\tan\,\delta \propto \left(1+\frac{N}{N_C}\right)^{-\phi}$ with $\phi<0.5$ was needed \cite{interactingTLS}. Here we fit our measurements to the phenomenological power-law and to two other alternative models: the interacting TLS model \cite{interactingTLS} and a model which assumes two types of TLSs \cite{SchechterTwoType}. 

\begin{figure}[h!]
	\subfloat[]{
		\includegraphics[width=0.5\linewidth]{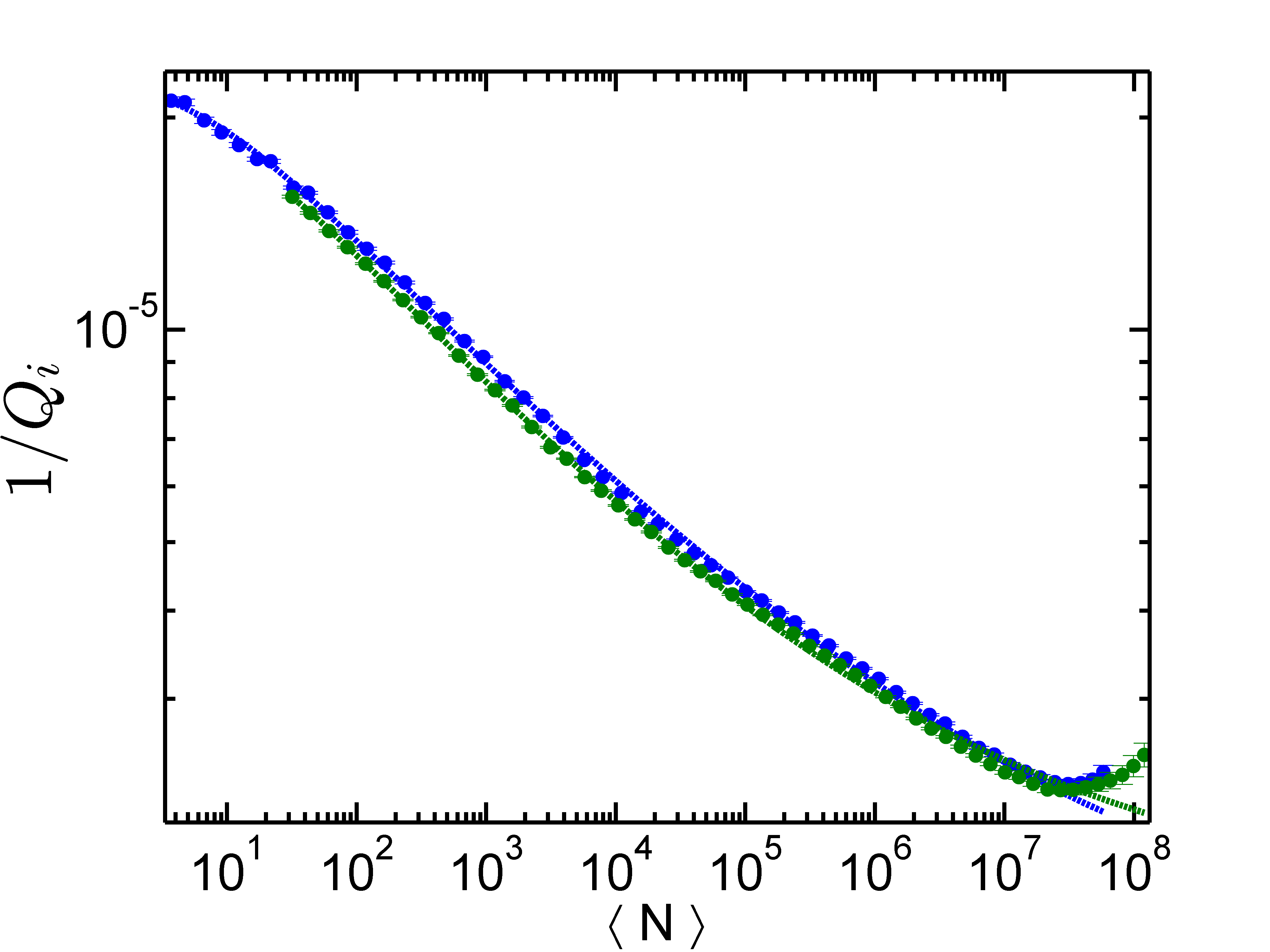} 
		\label{fig:standardLoss}
	}
	\hfill
	\subfloat[]{
		\includegraphics[width=0.5\linewidth]{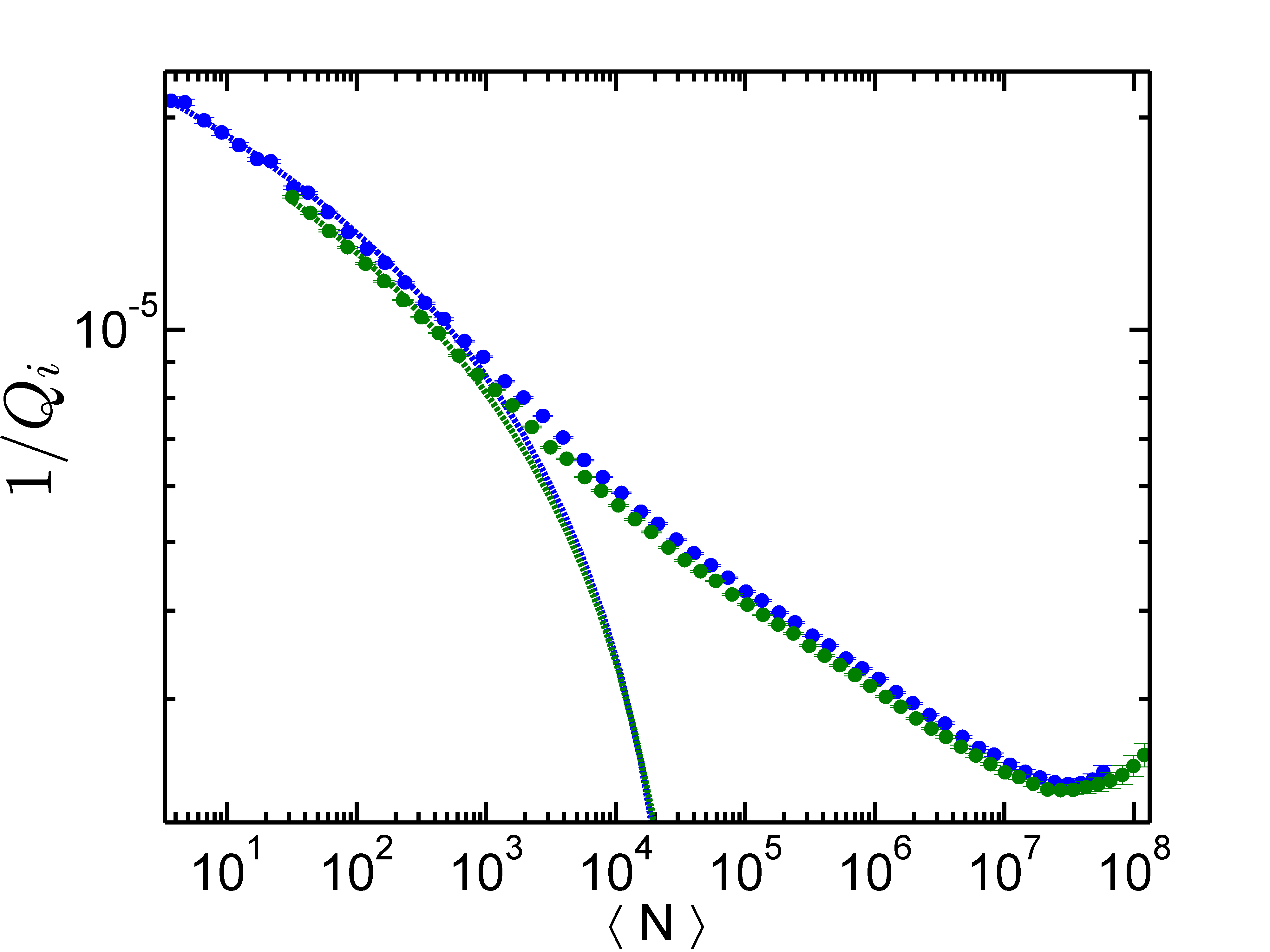} 
		\label{fig:logLoss}
	}
	\hfill
	\subfloat[]{
		\includegraphics[width=0.5\linewidth]{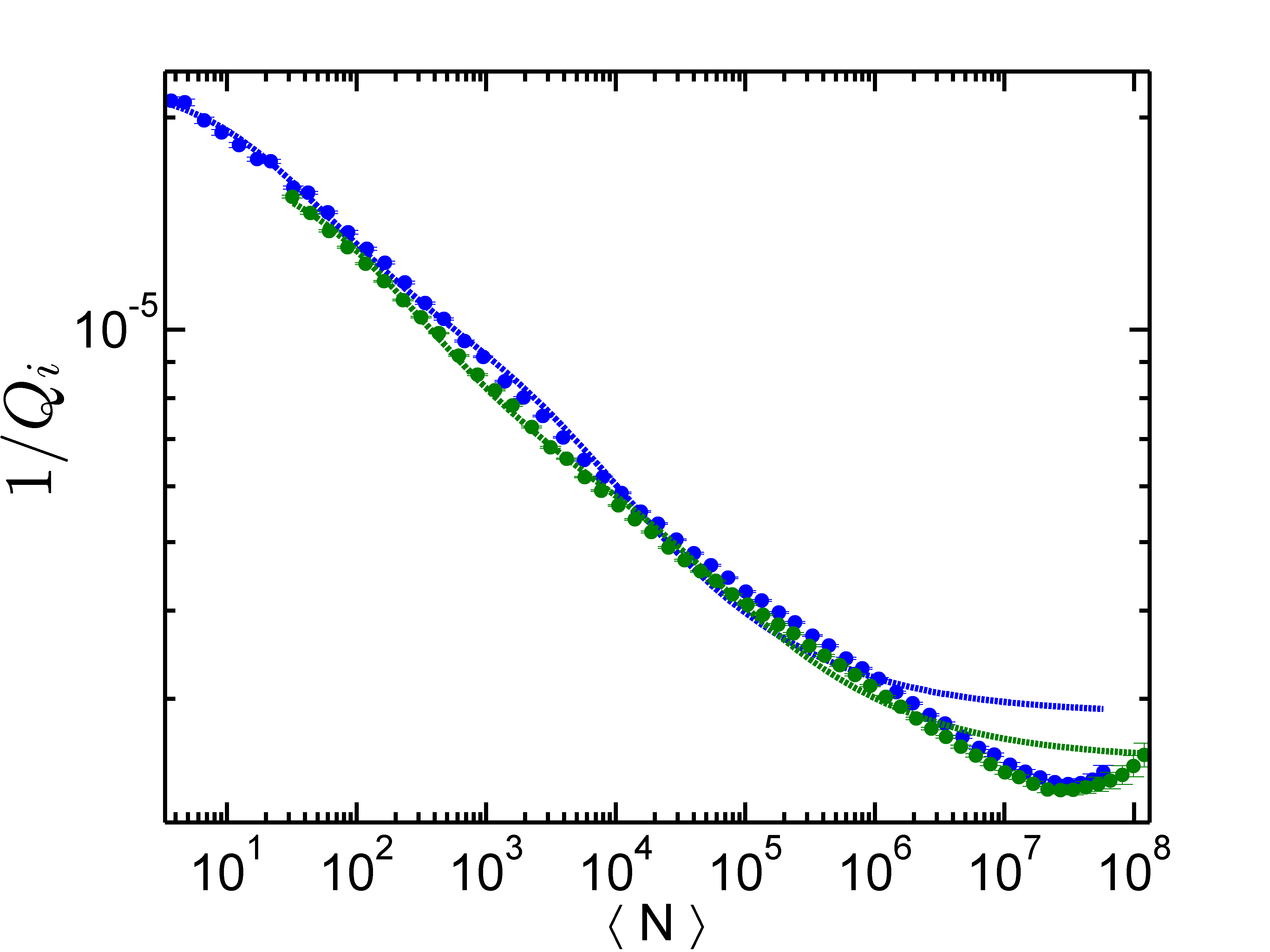} 
		\label{fig:twoTypesLoss}
	}
	\caption{Fitting the internal loss vs $\langle N \rangle$ for the probe-only experiments to the (a) Phenomenological power-law (b) interacting and (c) two TLS types models.  
		Green (blue) points are for the resonance at 5626 MHz (5689 MHz). Lines are fits to the various models.}
\end{figure}

Fig.~\ref{fig:standardLoss} shows fits to the phenomenological power-law
\begin{equation}
	\tan{\delta} = \tan{\delta_i}\cdot\left(1+\frac{N}{N_C}\right)^{-\phi}+\tan\,\delta_r,
	\label{eq:standardLossPheno}
\end{equation}
where $\tan\,\delta_r$ is a residual power-independent loss term. High photon-number measurements ($ \langle N \rangle > 10^7 $) in which the loss increases with power \cite{ResonatorsReview} were excluded from the fits. The extracted parameters are $ \phi \approx 0.21$, $ N_C \approx 7 $, $ \tan{\delta_i} \approx 2.1\times 10^{-5}$ and $ \tan\,\delta_r \approx 1.3\times 10^{-6}$. While this model fits the results, the ad-hoc exponent  $\phi \approx 0.21$ which is different than the theoretical $\phi=0.5$ and the seemingly unphysical low critical photon number $ N_C $ makes this model unfavourable. 

In Fig.~\ref{fig:logLoss} we show fits of the internal loss to the logarithmic dependence predicted by the interacting TLS model \cite{interactingTLS,BurnettIneractingTLS}
\begin{equation}
		\tan{\delta} = P_\gamma\tan{\delta_i}\log\left({\frac{\gamma_{max}}{\Omega_0}}
		\right)=P_\gamma\tan{\delta_i}\log\left({\frac{\gamma_{max}}{\Omega_0^{N=1}\sqrt{N}}}
		\right)=\frac{1}{2}P_\gamma\tan{\delta_i}\log\left({\frac{\left(\gamma_{max}/\Omega_0^{N=1}\right)^2}{N}}\right),
\end{equation}
where $ \gamma_{max} $ ($\gamma_{min}$) is the maximum (minimum) switching rate of fluctuating TLSs coupled to a coherent TLS, $P_\gamma\equiv\left[\log\left(\frac{\gamma_{max}}{\gamma_{min}}\right)\right]^{-1}$ is a normalization constant related to averaging over the switching rates and $\Omega_0$ ($\Omega_0^{N=1}$) is the TLSs (single-photon) Rabi frequency.    
Since this model is relevant only for low powers (where the fluctuations due to the interactions are faster than the Rabi frequency) we have fitted this model only for $\langle N \rangle < 1000$. This model also seem to fit the low power results. The extracted parameters yield (using the low-power value of $\tan{\delta}$ as $\tan{\delta_i}$) $P_\gamma \approx 0.2$ which is in the order of the expected value assuming switching rates in the range $\gamma \approx 1-10^6$ Hz \cite{BurnettIneractingTLS}. 
In addition, we obtain  $\frac{\gamma_{max}}{\Omega_0^{N=1}} \approx 220$ which for our results of $\Omega_0^{N=1}/2\pi \approx 80$ kHz gives $\gamma_{max} \approx 18$ MHz, which is reasonably self-consistent with the $P_\gamma$ value. We notice that this value of $\gamma_{max}$ cannot be explained as resulting from phonon mediated relaxation, since for thermal fluctuators with energy $E\equiv\hbar\omega\sim k_{B} T$ the maximal phonon relaxation rate given by Eq.~(\ref{eq:TLSreldec}) is $T_{1,min}^{-1} \sim AT^3$ which for 20 mK is $T_{1,min}^{-1}  \sim 1$ kHz.

Finally, the loss measurements were also fitted to a two-typed TLS model \cite{SchechterTwoType} 
\begin{equation}
\tan{\delta} =  \tan{\delta_{i1}}\left(1+\frac{N}{N_{C1}}\right)^{-0.5}+\tan{\delta_{i2}}\left(1+\frac{N}{N_{C2}}\right)^{-0.5}+\tan{\delta_r},
\end{equation}
where each TLS type $j$ has a different critical photon number $ N_{Cj} $ and an intrinsic low-power loss $\tan{\delta_{ij}}$ which is related to its coupling to the electric field, density of states and filling factor. The fits are shown in Fig.~\ref{fig:twoTypesLoss}. High photon-number measurements ($ \langle N \rangle > 10^7 $) in which the loss increases with power were excluded from the fit. The fits seem reasonable for $ \langle N \rangle < 10^4 $ but deviate for higher powers. The extracted parameters vary significantly for the two different resonances and are $ \tan{\delta_{i1}} \approx 5.2\times10^{-6},\, N_{C1} \approx 3300,\, \tan{\delta_{i2}} \approx 1.4\times10^{-5},\, N_{C2}\approx 15 $ and $ \tan{\delta_r} \approx 2.9\times10^{-6} $ for the resonance at 5689 MHz and $  \tan{\delta_{i1}} \approx 2.7\times10^{-6},\, N_{C1} \approx 27000,\, \tan{\delta_{i2}} \approx 1.2\times10^{-5},\, N_{C2}\approx 70 $ and $ \tan{\delta_r} \approx 2.5\times10^{-6} $ for the resonance at 5626 MHz. Although the actual values of the critical photon number differ significantly between both resonances their ratio is of the same order $\frac{N_{C1}}{N_{C2}} \approx 2-4\times 10^2$. This agrees with the assumption of the model that one TLS type has a much stronger coupling to strain and hence a shorter $T_1$ yielding a larger $N_C$ with $\frac{N_{C1}}{N_{C2}} \approx 10^2$ \cite{TLS_N_ratio}. The fact that $\tan{\delta_{i1}}<\tan{\delta_{i2}}$ is explained by the assumption that the strongly coupled TLSs are rare (i.e. have a smaller filling factor).     
We note that a two-typed TLS model was used to fit loss measurements in atomic layer deposition oxides \cite{twoTypesOsborn} and that two types of TLSs with a different dipole moment magnitude were found when individual TLSs were measured \cite{twoTypesOsborn2} and by echo measurements \cite{Echo}. 
In addition, two-typed TLS model was used to explain the strain dependence of echo dephasing in a recent experiment \cite{matityahu2016decoherence,lisenfeld2016decoherence}.       

To conclude, our loss measurements cannot distinguish between the various TLS models, but seem to be fitted well by models other than the phenomenological power-law. Further discussion regarding these competing models \cite{burin2015low,BurnettIneractingTLS} is outside the scope of this Letter. We notice that the dependence of the real part of the dielectric constant $ \epsilon $ on power does not change when interaction between TLSs are introduced \cite{interactingTLS}, justifying our ignoring of TLSs interactions when calculating the frequency shift.     

\section{Estimating the Kerr coefficient from nonlinear kinetic inductance}
\label{sec:KerrEstimation}
In this section we derive an order of magnitude estimation for the Kerr coefficient based on nonlinear kinetic inductance calculations and compare it to our results. Assuming the total (current dependent) inductance to be in the form \cite{ResonatorsReview} $ L = L_0+L_k\left[ 1+\left(\frac{I}{I_c}\right)^2\right]$ where $ L_0 $ and $ L_k $ are the geometric and (linear) kinetic inductances respectively and $ I_c $ is the critical (pair-breaking) current, we obtain the nonlinear frequency shift
\begin{equation}
	\Delta f\left(I\right) \approx -\frac{1}{2\pi} \frac{\omega_0^2}{2} \frac{L_k}{L_0} \left(\frac{I}{I_c}\right)^2. 
\end{equation}
Substituting the average number of photons $ \hbar\omega_0\langle N \rangle = LI^2$ we arrive at the expression for the Kerr coefficient 
\begin{equation}
	\frac{K}{2\pi}  \approx -\frac{1}{2\pi} \frac{\hbar\omega_0^2}{2L_0} \frac{L_k}{L_0} \left(\frac{1}{I_c}\right)^2.
\end{equation}  
For our devices $ L_0 \approx 1 $ nH and $ \omega_0 \approx 2\pi\times 5.6$ GHz. Using the values $ \frac{L_k}{L_0} = 0.04 $ for the kinetic inductance fraction \cite{GaoKIfrac} and  $ J_c = 100\,GA/m^2$ for the critical current density \cite{criticalCurrent} with a cross section of $ A = 120\,nm \times 8\,\mu m $ we obtain $ \frac{K}{2\pi} \approx -4.5\times10^{-5}$, which is in the order of magnitude of the values extracted from our measurements. 


\clearpage
	
%